\documentclass[aps,prb,showkeys,showpacs,superscriptaddress,floatfix,twocolumn]{revtex4-1}%
\usepackage{amsmath}
\usepackage{amsfonts}
\usepackage{amssymb}
\usepackage{graphicx}
\usepackage{etoolbox}
\apptocmd{\sloppy}{\hbadness 10000\relax}{}{}

\begin{document}

\title{Ultrafast supercontinuum generation in bulk condensed media (Invited~Review)}



\author{Audrius Dubietis}
\email{Corresponding author. e-mail: audrius.dubietis@ff.vu.lt}
\author{Gintaras Tamo\v{s}auskas}
\author{Rosvaldas \v{S}uminas}
\affiliation{Department of Quantum Electronics, Vilnius University, Saul\.{e}tekio Avenue 10, LT-10223 Vilnius, Lithuania}
\author{Vytautas Jukna}
\affiliation{Laboratoire d'Optique Appliqu\'{e}e, ENSTA ParisTech, Ecole Polytechnique, Universit\'{e} Paris-Saclay, F-91762 Palaiseau, France}
\affiliation{Centre de Physique Th\'{e}orique, CNRS, Ecole Polytechnique, Universit\'{e} Paris-Saclay, F-91128 Palaiseau, France} 
\author{Arnaud Couairon}
\affiliation{Centre de Physique Th\'{e}orique, CNRS, Ecole Polytechnique, Universit\'{e} Paris-Saclay, F-91128 Palaiseau, France}

\begin{abstract}
Nonlinear propagation of intense femtosecond laser pulses in bulk
transparent media leads to a specific propagation regime, termed
femtosecond filamentation, which in turn produces dramatic
spectral broadening, or superbroadening, termed supercontinuum
generation. Femtosecond supercontinuum generation in transparent
solids represents a compact, efficient and alignment-insensitive
technique for generation of coherent broadband radiation at
various parts of the optical spectrum, which finds numerous
applications in diverse fields of modern ultrafast science. During
recent years, this research field has reached a high level of
maturity, both in understanding of the underlying physics and in
achievement of exciting practical results. In this paper we
overview the state of the art of femtosecond supercontinuum
generation in various transparent solid-state media, ranging from
wide-bandgap dielectrics to semiconductor materials and in various
parts of the optical spectrum, from the ultraviolet to the
mid-infrared. A particular emphasis is given to the most recent
experimental developments: multioctave supercontinuum generation
with pumping in the mid-infrared spectral range, spectral control,
power and energy scaling of broadband radiation and the
development of simple, flexible and robust pulse compression
techniques, which deliver few optical cycle pulses and which could
be readily implemented in a variety of modern ultrafast laser
systems.
\end{abstract}

\pacs{42.65.Jx, 42.65.Re}

\keywords{Supercontinuum generation; femtosecond filamentation;
optical parametric amplification; pulse compression}

\maketitle

\section{Introduction}

One of the most spectacular and visually perceptible effects
produced by the nonlinear propagation of intense femtosecond laser
pulses in a transparent medium is dramatic broadening of the
spectrum, termed white-light continuum or supercontinuum (SC)
generation. At present, femtosecond SC represents a unique and
versatile source of coherent radiation, whose wavelength range
spans over a considerable part of the optical spectrum and by far
exceeds the definition of the white light.

During the last decade or so, a remarkable progress has been
achieved in femtosecond SC generation in photonic crystal
\cite{Dudley2006} and microstructured \cite{Price2012} fibers.
Technological availability of highly nonlinear materials, such as
tellurite and chalcogenide glasses, fostered the development of
ultrabroadband fiber-based SC sources, which deliver continuous
spectral coverage from the near to mid-infrared, see e.g.
\cite{Domachuk2008,Petersen2014}. Thanks to the development of
novel high peak power laser systems based on optical parametric
chirped pulse amplification (OPCPA) which operate in the
mid-infrared, a series of outstanding results on high power SC
generation in atmospheric air and other gases have been produced
recently \cite{Kartashov2012,Mitrofanov2015}.

SC generation in bulk condensed media deserves a special place in
the field due to its historical, scientific and technological
importance. The first observation of the white-light continuum in
a bulk solid state medium dates back to the early years of
nonlinear optics, when Alfano and Shapiro discovered spectral
broadening throughout the visible spectral range when focusing
powerful picosecond pulses into borosilicate glass sample
\cite{Alfano1970a}. The discovery was immediately followed by the
observations of spectral broadening in various crystals and
glasses, confirming the universal nature of the phenomenon
\cite{Alfano1970b}, also see \cite{Alfano2006} for a complete
historical account on the early developments of SC generation in
various optical media.

The first femtosecond SC was generated in 1983, when Fork and
co-authors reported spectral broadening from the deep ultraviolet
to the near infrared by focusing intense 80-fs pulses at 627 nm
from the dye laser into an ethylene glycol jet \cite{Fork1983}.
This study also underlined new qualities of broadband radiation,
such as improvement of pulse-to-pulse reproducibility and spatial
uniformity, which benefited from the short duration of the input
pulse. Since then femtosecond supercontinuum generation using dye
driving lasers was demonstrated in various condensed bulk media,
see \cite{Wittmann1996} and references therein.

The discovery of Kerr lens mode locking, which led to the
invention of femtosecond Ti:sapphire laser oscillator
\cite{Spence1991}, which was followed by the demonstration of
regenerative amplification of the oscillator pulses, marked a new
era in femtosecond SC generation \cite{Norris1992}. The amplified
Ti:sapphire lasers outperformed then widely spread femtosecond dye
lasers in all essential parameters of operation, setting a new
standard for the generation of high power femtosecond pulses
\cite{Backus1998}. The advances in femtosecond solid-state laser
technology as combined with growing practical knowledge of
femtosecond SC generation in transparent condensed media
\cite{Brodeur1998,Brodeur1999}, boosted the development of
femtosecond optical parametric amplifiers \cite{Wilson1997}, which
made ultrashort broadly tunable pulses routinely available. These
developments in turn facilitated the experimental studies of the
SC generation in previously poorly explored mid-infrared spectral
domain, provided an access to the anomalous group velocity
dispersion (GVD) region of transparent media, see e.g.
\cite{Couairon2016} and markedly extended the nomenclature of
suitable nonlinear materials.

In practice, femtosecond SC generation in condensed bulk media
constitutes a compact, efficient, low cost, highly robust and
alignment-insensitive technique for the generation of coherent
ultrabroadband radiation at various parts of the optical spectrum.
As being induced by self-focusing and filamentation of intense
femtosecond laser pulses, the SC radiation bears high spatial and
temporal coherence \cite{Chin1999}. The SC beam preserves the same
level of spatial coherence as the input laser beam
\cite{Watanabe2001}, or even shows its improvement as due to
spatial mode cleaning, which is a universal feature of the
filamentation process \cite{Prade2006}, yielding an excellent
focusability of the SC beam. High temporal coherence of the SC
radiation stems from a well-defined temporal structure of
femtosecond filament. The SC pulses acquire a regular chirp due to
material dispersion and hence exhibit good compressibility, as
verified by spectral phase characterization and post-compression
experiments \cite{Wegkamp2011}. In general, the SC radiation
produced by a single filament maintains the polarization state of
the input pulses; the depolarization effects start to manifest
themselves only at large input powers, well exceeding the critical
power for self-focusing
\cite{Midorikawa2002,Dharmadhikari2006c,Kumar2008}. The SC
radiation has reasonable spectral density ($\sim10$ pJ/nm)
\cite{Bradler2009} and exhibits supreme statistical properties,
such as mutual correlations between the intensities of the
spectral components, low pulse-to-pulse fluctuations, and
excellent long-term stability, which compares to that of the pump
laser source itself
\cite{Bradler2009,Megerle2009,Majus2011,Majus2013,Bradler2014,Walle2015}.

Altogether, these outstanding properties make femtosecond SC
highly suitable for diverse applications in ultrafast
spectroscopy, photonics, femtosecond technology and contemporary
nonlinear optics. Extremely broad SC spectrum is on demand for
transient pump-probe spectroscopy and gives an access to study
ultrafast molecular dynamics and processes in condensed matter
with femtosecond time resolution, see e.g.
\cite{Megerle2009,Calabrese2012,Aubock2012,Riedle2013}. The use of
SC for Z-scan measurements enables broadband characterization of
the nonlinear absorption spectra \cite{Balu2004,DeBoni2004}.
Multiple filaments emerging from self-focusing of a single high
power input beam exhibit mutual coherence and provide a source of
multichannel white-light radiation \cite{Cook2003,Corsi2004}.
Independently generated phase-locked SC pulses
\cite{Bellini2000,Bellini2001,Tortora2004} as well as
parametrically amplified portions of independent SC spectra
\cite{Baum2003,Baum2005} serve for production of frequency combs
allowing for absolute high precision frequency measurements within
a broad spectral range. The specific SC emission patterns, termed
``ciliary white light'', which emerge from the damage craters in
transparent dielectric media, carry rich information on the damage
profile and morphology dynamics of the ablated surface, providing
a real time in situ observation of the laser ablation process
\cite{Liu2013}. Most importantly, SC is recognized as an
indispensable source for seeding ultrafast optical parametric
amplifiers, which provide broadly tunable femtosecond pulses
\cite{Cerullo2003}. Broad spectral bandwidth and compressibility
of the SC pulses served as an important asset that contributed to
the invention of ultrabroadband noncollinear optical parametric
amplifiers \cite{Wilhelm1997}, which currently produce few optical
cycle pulses at various parts of the optical spectrum, ranging
from the visible to the mid-infrared \cite{Brida2010}. Moreover,
the SC pulses preserve the carrier envelope phase of the pump
pulses \cite{Manzoni2009}, whose stability is of key importance in
determining the interaction of few optical cycle pulses with
matter. The SC seeding has set new landmarks in simplifying the
architecture of table-top OPCPA technique-based laser systems
\cite{Dubietis2006}. Recently developed high repetition rate
SC-seeded OPCPA systems provide intense few optical cycle pulses
with high average power at high repetition rate \cite{Harth2012}.
SC generation in laser host crystals with sub-picosecond and
picosecond laser pulses paved a new avenue in the development of
compact OPCPA systems, which are built around sole picosecond
oscillator/amplifier systems \cite{Schulz2011}. Finally, SC-seeded
OPCPA system delivering carrier envelope phase stable few optical
cycle pulses with high average and ultrahigh peak power was
recently demonstrated \cite{Budriunas2017}.

In this Review we present a comprehensive outlook of the results
on femtosecond SC generation in transparent condensed media, which
have been achieved over the last 15 years or so. The paper is
organized as follows: in section 2 we present the underlying
physical picture of SC generation, which is based on femtosecond
filamentation, emphasizing the temporal dynamics of the SC
generation and the role of material dispersion, in particular; in
section 3 we briefly discuss the governing equations and numerical
models for nonlinear propagation of intense femtosecond laser
pulses in transparent dielectric media; in section 4 we provide
the list of wide bandgap dielectric materials and their optical
properties, which are relevant for SC generation, and useful
experimental issues, which are of importance for optimizing the
practical schemes of SC generation in bulk dielectric media; in
section 5 we present an overview of experimental results in
commonly used wide-bandgap dielectric materials: silica and
non-silica glasses, water, alkali metal fluorides, laser hosts and
birefingent nonlinear crystals with second-order nonlinearity;
section 6 is devoted to SC generation in semiconductor crystals,
which hold a great potential in the rapidly developing field of
ultrafast mid-infrared nonlinear optics; section 7 presents a
brief summary of the experimental results aiming at SC generation
with controlled spectral extent; in section 8 we discuss the
recent developments in power scaling of the SC and pulse
compression techniques based on spectral broadening; finally in
section 9 we briefly discuss other developments in the field,
which include generation of odd harmonics-enhanced SC, SC
generation in narrow bandgap dielectric media, and SC generation
with Bessel and Airy beams and optical vortices.

\section{Physical picture of supercontinuum generation}

In contrast to optical fibers, where the propagation dynamics of
the optical pulse is essentially one-dimensional and the spectral
broadening arises from the soliton generation and fission as due
to the interplay between the self-phase modulation and the
material dispersion, see e.g. \cite{Dudley2006}, the SC generation
in bulk media appears to be a more complex process that involves
an intricate coupling between spatial and temporal effects. The
physical picture of SC generation in transparent bulk media could
be understood in the framework of femtosecond filamentation, which
provides a universal scenario of nonlinear propagation and
spectral broadening of intense femtosecond laser pulses in solids,
liquids and gases
\cite{Chin2005,Couairon2007,Berge2007,Kandidov2009}.

Femtosecond filamentation stems from the interplay between
self-focusing, self-phase modulation, multiphoton
absorption/ionization-induced free electron plasma, leading to the
appearance of ``a dynamic structure with an intense core, that is
able to propagate over extended distances much larger than the
typical diffraction length while keeping a narrow beam size
without the help of any external guiding mechanism'', termed
femtosecond filament \cite{Couairon2007}. Generation of an
extremely broadband, spatially and temporally coherent emission
with a low angular divergence (or SC generation) is the most
obvious manifestation of filament formation. The SC emission is
accompanied by the generation of colored conical emission, i.e.
broadband radiation that is emitted at different angles with
respect to the propagation axis, forming a beautiful array of
concentric colored rings. In the case of condensed media (solids
and liquids) these effects are brought to a compact, few
millimeter to few centimeter length scale.

\begin{figure}[ht!]
\includegraphics[width=8.5cm]{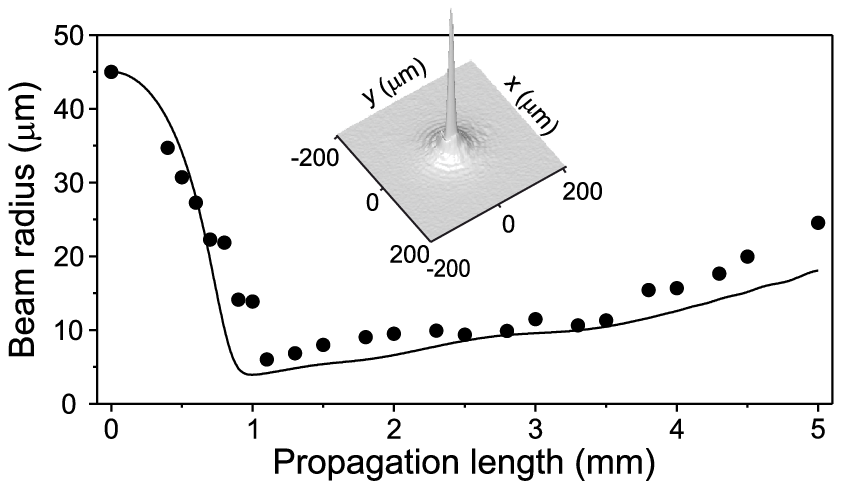}
\caption{FWHM beam radius as a function of propagation distance
during self-focusing of a loosely focused Gaussian input beam with
$90~\mu$m FWHM diameter in water. The input beam power is
$\sim5~P_{\rm cr}$. Solid curve shows the numerical simulation,
bold dots show the experimental data. The inset shows fluence
profile of the filament, which forms after the nonlinear focus.
Adapted from \cite{Dubietis2006a}.}\label{fig:selffoc}
\end{figure}

The initial stage of filament formation is a result of the
intensity-dependent refractive index: $n=n_0+n_2I$, where $I$ is
the intensity, $n_0$ is the linear refractive index. $n_2$ is the
nonlinear refractive index, which is related to the third-order
(cubic) optical susceptibility of the material and which is
positive in the transparency range of dielectric media. The
induced change of the refractive index is proportional to the
local intensity and thus is higher at the center of the beam and
lower at the edges. Therefore the material acts like a lens, which
enforces the beam to self-focus. For a cylindrically symmetric
Gaussian beam the self-focusing threshold is defined by the beam
power

\begin{equation}
P_{\rm cr}=\frac{3.72\lambda^2}{8\pi n_0n_2}.
\label{eq:pcr}\end{equation}

where $\lambda$ is the laser wavelength, and which is called the
critical power for self-focusing; that is the power, when the
effect of self-focusing precisely balances the diffractive
spreading of the beam. This balance is realized for the so-called
Townes beam, whose shape is close to a Gaussian beam. If the beam
power exceeds $P_{\rm cr}$, the collimated input Gaussian beam
will self-focus at a distance \cite{Marburger1975}:

\begin{equation}
z_{\rm sf}=\frac{0.367z_R}{\sqrt{[(P/P_{\rm
cr})^{1/2}-0.852]^2-0.0219}}. \label{margurger}
\end{equation}

which is called the nonlinear focus. Here $z_R=\pi
n_0w_0^2/\lambda$ denotes the Rayleigh (diffraction) length of the
input Gaussian beam of a radius $w_0$. Although
Eq.~(\ref{margurger}) is derived in the case of continuous wave
laser beams, it gives a fairly accurate approximation of the
nonlinear focus of femtosecond laser pulses as well.
Figure~\ref{fig:selffoc} shows an example of the evolution of the
beam radius during self-focusing of a loosely focused
femtosecond-pulsed Gaussian beam with power of $\sim5~P_{\rm cr}$
in water. The position of the nonlinear focus is indicated by the
minimum beam radius.

In the time domain, a time-varying refractive index imparts a
nonlinear change in the phase of the pulse

\begin{equation}
\phi_{\rm nl}(t)=-\frac{\omega_0}{c}n_2I(t)z,
\end{equation}

where $\omega_0$ is the carrier frequency and $z$ is the
propagation distance, which produces a frequency change
$\delta\omega(t)=\frac{d}{dt}\phi_{\rm nl}(t)$ that results in a
time-varying instantaneous frequency:

\begin{equation}
\omega(t)=\omega_0+\delta\omega(t),
\end{equation}

giving rise to spectral broadening of the pulse. For a Gaussian
laser pulse with the pulse duration $t_p$, variation of the
instantaneous frequency is expressed as

\begin{equation}
\delta\omega(t)=-2\frac{\omega_0}{ct_p^2}n_2I_0\exp\left(-\frac{t^2}{t_p^2}\right)tz.
\end{equation}

\begin{figure}[ht!]
\includegraphics[width=8.5cm]{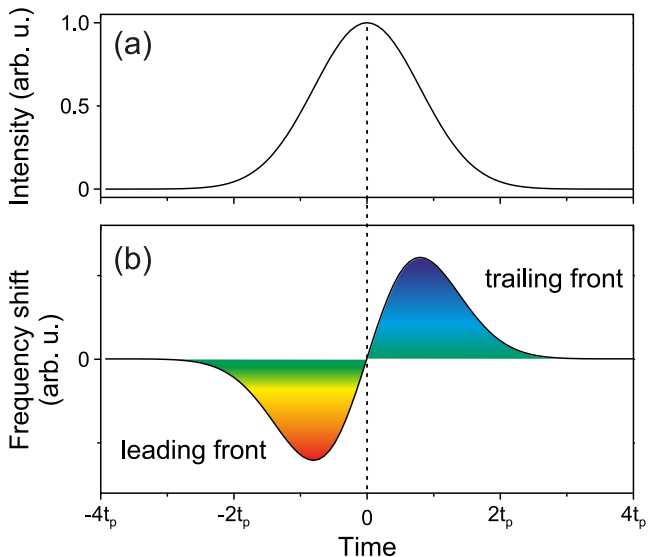}
\caption{Self phase modulation of a Gaussian pulse shown in (a),
which produces a variation of the instantaneous frequency shown in
(b).}\label{fig:spm}
\end{figure}

The effect is called self-phase modulation and gives rise to
spectral broadening by inducing a negative shift of the
instantaneous frequency at the leading (ascending) front of the
pulse and a positive shift of the instantaneous frequency at the
trailing (descending) front of the pulse, as schematically
illustrated in Fig.~\ref{fig:spm}. In other words, the pulse
acquires a frequency modulation corresponding to the production of
red-shifted spectral components at the pulse front and the
blue-shifted spectral components at the pulse tail.

The self-focusing stage is a runaway effect in the sense that as
the beam self-focuses, the intensity increases and so does the
self-focusing effect. However, the beam cannot focus to a
singularity; the beam collapse at the nonlinear focus is arrested
by the multiphoton absorption and ionization, producing an energy
loss and generating a free electron plasma, which further absorbs
and defocuses the beam. A combined action of these effects limits
or clamps the intensity to a certain level. The clamping intensity
depends on the order of multiphoton absorption: $K=\langle
U_g/\hbar\omega_0\rangle+1$, where $U_g$ is the bandgap,
$\hbar\omega_0$ is the photon energy
\cite{Brodeur1998,Brodeur1999}. The higher is the order of
multiphoton absorption, the higher is clamping intensity and the
smaller is the limiting beam diameter at the nonlinear focus, so
the larger spectral broadening is produced.

These simple considerations provide a plausible explanation of the
experimentally observed bandgap dependence of the SC spectral
extent and suggest that the broadest SC spectra could be attained
in wide bandgap dielectrics \cite{Brodeur1998,Brodeur1999}. In
contrast, self-focusing in the case of $K<3$ cannot produce SC.
However, it is interesting to note the inverse relationship
between $E_g$ and $n_2$; the larger is the bandgap, the smaller is
the value of $n_2$ \cite{Sheik-Bahae1990}. This is quite a
paradox, since $n_2$ defines the strength of self-focusing and
self phase modulation, which are the fundamental physical effects
behind femtosecond filamentation.

The intensity clamping effect was verified experimentally by
measuring a constant width of the SC spectrum for a wide range of
pulse energies above a threshold input laser energy for SC
generation \cite{Liu2002a}. In condensed media the maximum clamped
intensities up to tens of TW/cm$^2$ inside the nonlinear medium
are estimated, accounting only for the balance between
self-focusing and defocusing by free electron plasma generated by
multiphoton absorption. However, plasma defocusing and absorption
become relevant for the input pulse intensities of few TW/cm$^2$,
contributing to significant shortening of the pulses before the
nonlinear focus. If the input beam energy and focusing conditions
are properly chosen, the catastrophic avalanche ionization does
not come into play, so the plasma density is kept below the
critical value ($10^{21}$ cm$^{-3}$) and optical damage of the
material is avoided.

Numerical studies uncovered that besides the intensity clamping,
the chromatic dispersion is an equally important player, which
determines the extent and shape of the SC spectrum
\cite{Kolesik2003a,Kolesik2003b}. The role of chromatic dispersion
could be fairly evaluated in the framework of the effective three
wave mixing, which interprets SC generation as the emergence of
new frequency components due to scattering of the incident optical
field from the material perturbation via nonlinear polarization
\cite{Kolesik2003a,Kolesik2003b}. From a simple and practical
viewpoint, this approach suggests that lower chromatic dispersion
allows fulfilment of the phase matching condition for a broader
range of scattered spectral components, that is, supports larger
spectral broadening and vice versa.

\begin{figure*}[ht!]
\includegraphics[width=17cm]{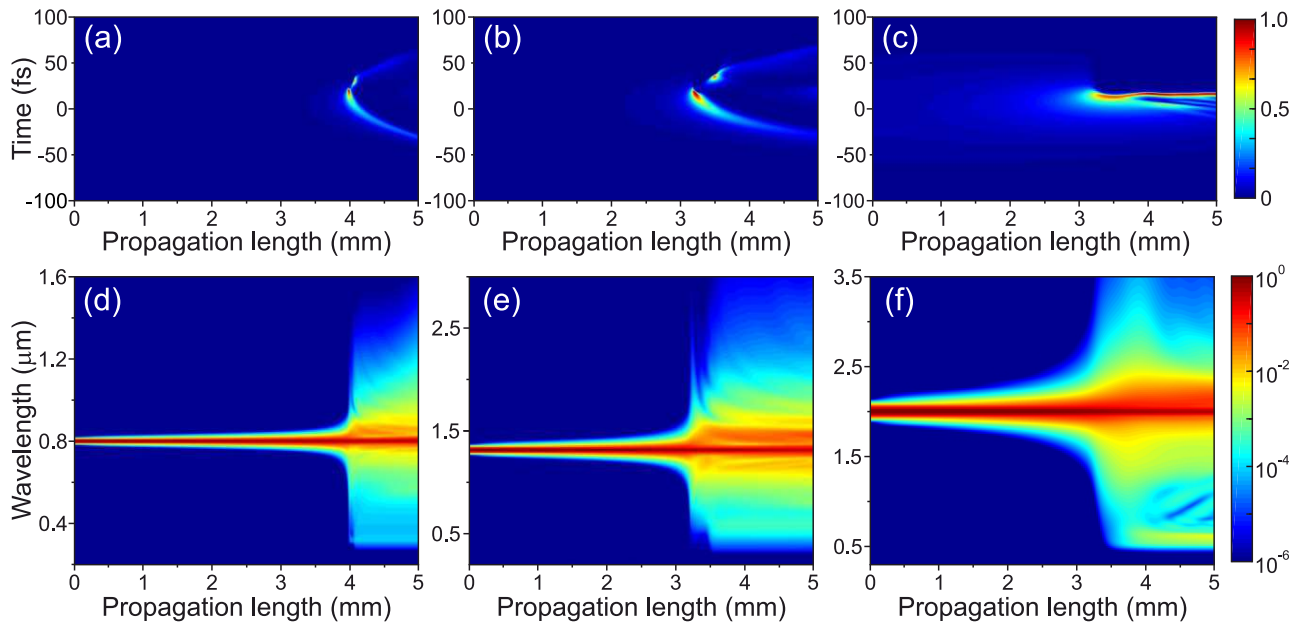}
\caption{Top row: numerically simulated temporal dynamics of 100
fs laser pulses propagating in sapphire crystal with the inputf
wavelengths of (a) 800 nm, (b) 1.3 $\mu$m, (c) 2.0 $\mu$m,
representing the filamentation regimes of normal, zero and
anomalous GVD, respectively. Bottom row shows the corresponding
spectral dynamics. Notice, how the spectral broadening in the
regimes of normal and zero GVD is associated with the pulse
splitting, and the spectral broadening in the regime of anomalous
GVD is associated with pulse self-compression.}\label{fig:time}
\end{figure*}

On the other hand, the interplay between the nonlinear effects and
chromatic dispersion gives rise to space-time focusing and
self-steepening, while the sign of GVD basically defines the
emerging temporal dynamics of femtosecond filament and so the
temporal and spectral content of the SC, see e.g.
\cite{Skupin2006} for an illustrative numerical study. The GVD
coefficient is defined as
$k''_0=\partial^2k/\partial\omega^2|_{\omega_0}$, where
$k=\omega_0n_0/c$ is the wavenumber. In the region of normal GVD
($k''_0>0$), the red-shifted frequencies travel faster than the
blue-shifted ones, while the opposite is true in the region of
anomalous GVD ($k''_0<0$). Figure~\ref{fig:time} compares the
numerically simulated temporal evolution of femtosecond filament
and respective spectral dynamics in sapphire crystal, in the
ranges of normal (the input wavelength 800 nm), zero ($1.3~\mu$m)
and anomalous ($2.0~\mu$m) GVD.

\subsection{Normal GVD}

As normal GVD of wide bandgap dielectric materials lies in the
wavelength range from the UV to the near IR, that is readily
accessible by modern femtosecond solid state lasers and their
harmonics, it is not surprising that the SC generation in that
spectral range has received the largest theoretical and
experimental attention.

In the region of normal GVD of dielectric media, pulse splitting
is the effect that governs the self-focusing dynamics and the
spectral broadening at and beyond the nonlinear focus. Pulse
splitting in normally dispersive dielectric medium has been
foreseen theoretically using the propagation models of different
complexity
\cite{Chernev1992,Rothenberg1992a,Rothenberg1992b,Fibich1997}. In
the input power range of just slightly above $P_{\rm cr}$, the
pulse splitting was proposed as the mechanism which contributes in
arresting the collapse of ultrashort pulses at the nonlinear
focus. The theoretical predictions were afterwards confirmed
experimentally by means of autocorrelation measurements
\cite{Ranka1996}. A more detailed information on pulse splitting
was extracted by recording the cross-correlation functions
confirming the numerically predicted asymmetry between the
intensities of the split sub-pulses \cite{Ranka1998}, while the
detailed amplitude structure and phase information of the split
sub-pulses was retrieved from the frequency resolved optical
gating (FROG) \cite{Diddams1998}, eventually establishing the
general link between pulse splitting and SC generation
\cite{Zozulya1999}.

These findings laid the basis of the pulse-splitting-based
temporal scenario of SC generation in normally dispersive media
\cite{Gaeta2000}. As the self-phase modulation broadens the pulse
spectrum and produces the nonlinear frequency modulation (chirp)
in which red-shifted and blue-shifted frequencies are generated at
the leading and trailing parts of the pulse, respectively, the
pulse splitting at the nonlinear focus produces two sub-pulses
with shifted carrier frequencies. Because of that, the sub-pulses
move in opposite directions in the frame of the input pulse, as
illustrated in Fig.~\ref{fig:time}(a). Pulse splitting is
immediately followed by an explosive broadening of the spectrum
[Fig.~\ref{fig:time}(a)], which is produced by the self-steepening
effect. The latter effect originates from velocity differences
between the pulse peak and the pulses tails as due to the
refractive index dependence on the intensity, and so induces sharp
intensity gradients (optical shocks) in the temporal profiles of
the sub-pulses.

The split sub-pulses experience rather different self-steepenings,
which quantitatively explain the asymmetry in experimentally
measured shapes of the SC spectra. In the near infrared spectral
range, under typical focusing conditions, a particularly steep
edge is formed at the trailing front of the trailing sub-pulse,
giving rise to a broad blue-shifted pedestal in the SC spectrum.
In contrast, the self-steepening of the leading front of the
leading sub-pulse is much less, resulting in rather modest
red-shifted spectral broadening. The connections between the
leading and trailing sub-pulses and the red-shifted and the
blue-shifted spectral broadenings, were verified experimentally by
measuring the spectral content of split sub-pulses
\cite{Gaeta2009}. The universality of pulse splitting based
scenario of SC generation in normally dispersive media is
confirmed by apparent similarity of the SC spectral shapes, as
generated in various nonlinear media.

\subsection{Anomalous GVD}

A qualitatively different temporal scenario of self-focusing and
femtosecond filamentation was foreseen in the range of anomalous
GVD, suggesting that the interplay between self-focusing,
self-phase modulation and anomalous GVD may lead to simultaneous
shrinking of the input wave packet in spatial and temporal
dimensions, giving rise to the formation of self-compressed three
dimensional (spatiotemporal) light bullets \cite{Silberberg1990}.
Here new red-shifted and blue-shifted frequencies, that are
generated by the self-phase modulation on the ascending (leading)
and descending (trailing) fronts of the pulse, respectively, are
swept back to the peak of the pulse, instead of being dispersed as
in the case of normal GVD. The feasibility of self-compressed
objects was confirmed by the numerical simulations using more
realistic numerical models \cite{Berge2005,Liu2006,Chekalin2013},
which predicted that pulse self-compression down to a single
optical cycle is potentially possible.

Development of high-peak-power near- and mid-infrared
ultrashort-pulse laser sources, which are exclusively based on the
optical parametric amplification gave an experimental access to
study filamentation phenomena in the range of anomalous GVD of
wide bandgap dielectrics and even semiconductors, whose zero GVD
wavelengths are located deeply in the mid-infrared. To this end, a
remarkably (almost 10 times) increased length of the filaments
\cite{Moll2004} and ultrabroadband SC emission
\cite{Saliminia2005} was observed by launching femtosecond pulses
at 1.55 $\mu$m in fused silica. A more recent study demonstrated
filamentation of incident pulses with much longer wavelength (3.1
$\mu$m) in YAG crystal, yielding more than 3 octave-spanning SC
spectrum with unprecedented wavelength coverage from the
ultraviolet to the mid-infrared \cite{Silva2012}. Eventually,
simultaneous time and space compression was demonstrated to favor
a new type of filamentation, which produces quasistationary
three-dimensional self-compressed light bullets that preserve a
narrow beam diameter and a short pulsewidth over a considerable
propagation distance in a nonlinear dispersive medium
\cite{Durand2013a}. To this end, the formation of self-compressed
spatiotemporal light bullets was experimentally observed in
various nonlinear media, such as fused silica, sapphire and BBO,
and under a variety of operating conditions
\cite{Smetanina2013a,Majus2014,Grazuleviciute2015,Chekalin2015,Suminas2016}.

Figure~\ref{fig:time}(c) shows a numerical example illustrating
formation and propagation dynamics of the self-compressed
spatiotemporal light bullet in sapphire crystal, which is
accompanied by generation of ultrabroadband SC
[Fig.~\ref{fig:time}(f)], which emerges at the point where the
pulse self-compression occurs.

\subsection{Zero GVD}

\begin{figure}[ht!]
\includegraphics[width=8.5cm]{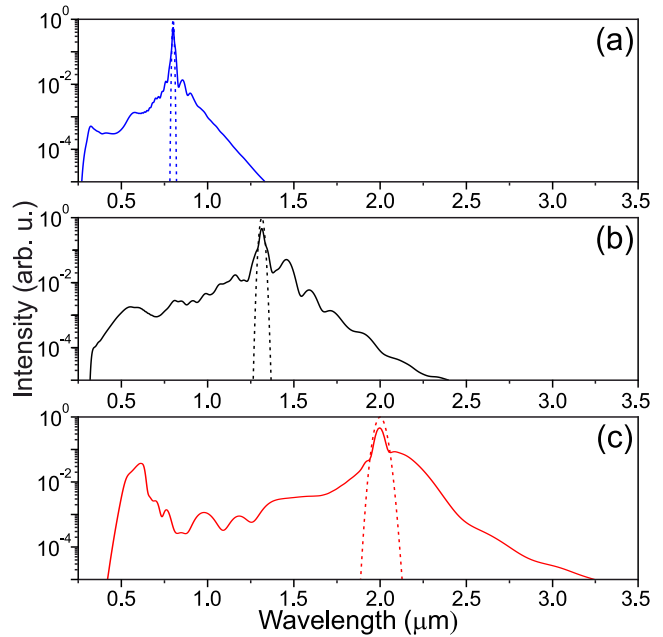}
\caption{Lineouts of the numerically simulated axial
supercontinuum spectra after 4 mm-thick sapphire crystal in the
cases of (a) normal GVD, (b) zero GVD, (c) anomalous GVD. The
input spectra are shown by the dashed curves.}\label{fig:num}
\end{figure}

The zero GVD wavelengths of commonly used wide bandgap dielectric
materials lie in the infrared spectral range, in the wavelength
interval between 1 and 2 $\mu$m, see Table~1. To some extent, the
near-zero GVD regime combines properties of both normal and
anomalous GVD, and produces more symmetric spectral broadening
than in the two previous propagation regimes \cite{Skupin2006}.
However, in the time domain, the input pulse undergoes splitting
at the nonlinear focus, as illustrated in Fig.~\ref{fig:time}(b),
and the post-collapse dynamics are essentially similar to those
observed in the case of normal GVD shown in
Fig.~\ref{fig:time}(a). Experimental measurements show that the
pulse splitting prevails even in the case of weak anomalous GVD
\cite{Grazuleviciute2016}, where the amount of material dispersion
is too small to compress the spectrally broadened pulse.

Figure~\ref{fig:num} presents a comparison of the numerically
simulated axial SC spectra generated by self-focusing and
filamentation of 100 fs pulses in 4 mm-thick sapphire crystal, in
the regimes of normal (the input wavelength 800 nm), zero (1.3
$\mu$m) and anomalous (2.0 $\mu$m) GVD.

\subsection{Conical emission}

\begin{figure}[ht!]
\includegraphics[width=8.5cm]{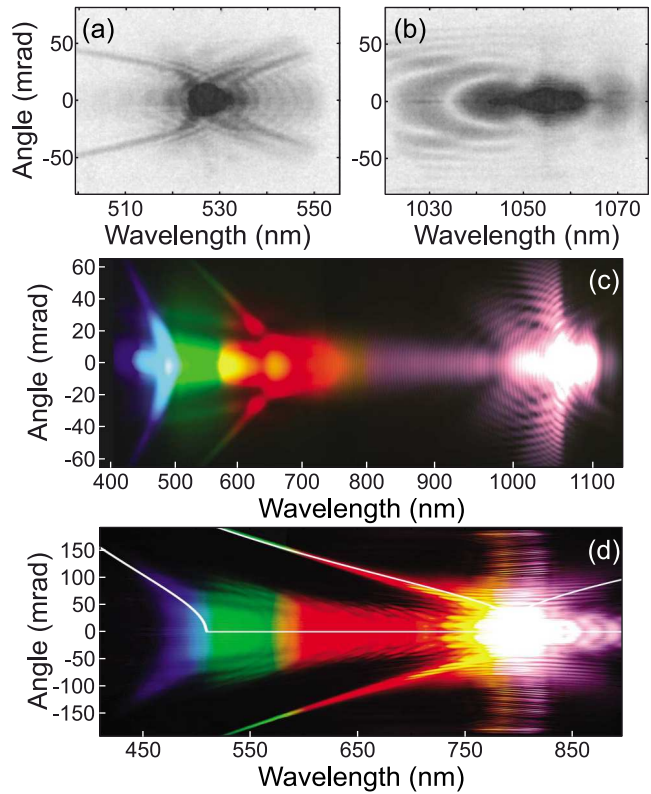}
\caption{Experimentally measured angle-resolved spectra around the
incident wavelengths of 527 nm and 1.055 $\mu$m that fall into the
ranges of (a) normal GVD, (b) anomalous GVD of water,
respectively. Adapted from \cite{Porras2005a}. (c) The entire
angle-resolved SC spectrum in water as excited with 1.055 $\mu$m
input pulses. Adapted from \cite{Faccio2008}. (d) The
angle-resolved SC spectrum as excited with 800 nm input pulses in
sapphire. The white solid curves indicate the best fits obtained
using the X-wave relation. Adapted from
\cite{Faccio2008a}.}\label{fig:ce}
\end{figure}

Finally, the above filamentation and SC generation scenarios could
be generalized by employing the effective three wave mixing model,
which provides the unified picture, connecting the spectral
broadening on the propagation axis with colored conical emission
\cite{Kolesik2003a,Kolesik2003b}, which is perhaps the most
striking and visually perceptible evidence of SC generation in
bulk media. The spectral content and the angular distribution of
the scattered waves satisfy phase matching conditions, which are
defined by the chromatic dispersion, thereby providing a
particular dispersion-defined angular landscape of scattered
frequencies \cite{Kolesik2005}. Experimentally, these landscapes
could be retrieved by measuring the SC spectrum with an imaging
spectrometer \cite{Faccio2005}. Figures~\ref{fig:ce}(a)-(c) show
the experimentally measured angularly resolved SC spectra in
water, which exhibit qualitatively different patterns of conical
emission in the range of normal and anomalous GVD
\cite{Porras2005a}. More specifically, in the range of normal GVD,
off-axis (conical) tails emerge on both the blue and red-shifted
sides of the input wavelength, forming a distinct X-shaped pattern
of conical emission, as shown in Fig.~\ref{fig:ce}(a). In
contrast, in the range of anomalous GVD, conical emission pattern
develops a multiple annular, or O-shaped structure around the
input wavelength, as illustrated in Fig.~\ref{fig:ce}(b). The
entire angle-resolved SC spectrum produced by filamentation of
1055 nm laser pulses, whose wavelengths falls into the range of
anomalous GVD of water, is presented in Fig.~\ref{fig:ce}(c). It
consists of multiple annular structures around the carrier
wavelength and a distinct V-shaped tail in the visible spectral
range \cite{Faccio2008}. Figure~\ref{fig:ce}(d) illustrates the
entire angle-resolved SC spectrum in sapphire, as generated by 800
nm pulses, in the range of normal GVD \cite{Faccio2008a}.

The shapes of angle-resolved SC spectra are universal for any
other nonlinear medium. These findings led to the interpretation
of femtosecond filaments as conical waves, assuming that the input
wave packet will try to evolve toward a final stationary state
that has the form of either an X-wave in the range of normal GVD
or an O-wave in the range of anomalous GVD. Nonlinear X-waves
\cite{Kolesik2004,Couairon2006} and O-waves \cite{Porras2005b} are
named because of their evident X-like and O-like shapes,
respectively, which appear in both the near and the far-fields.
Moreover, the interpretation of light filaments in the framework
of conical waves readily explains the distinctive propagation
features of light filaments such as sub-diffractive propagation in
free space \cite{Faccio2008a,Majus2014} and self-reconstruction
after hitting physical obstacles
\cite{Dubietis2004a,Dubietis2004b,Grazuleviciute2014}, which are
universal and regardless of the sign of material GVD, and which
were verified experimentally as well. Therefore all subsequent
features of the filament propagation in the regime of normal GVD,
i.e. pulse splitting, conical emission and any nonlinear
interactions, may be interpreted assuming the pulses as
spontaneously occurring nonlinear X-waves \cite{Faccio2006a}.
Consequently, the formation and propagation features of
spatiotemporal light bullets in the regime of anomalous GVD, may
be interpreted in terms of nonlinear O-waves \cite{Majus2014}.

\section{Numerical models}

A standard numerical model for nonlinear propagation and
filamentation of intense femtosecond laser pulse in transparent
dielectric medium with cubic nonlinearity is based on solving the
paraxial unidirectional propagation equation with cylindrical
symmetry for the nonlinear pulse envelope coupled with rate
equations describing the generation of free electrons by optical
field ionization. In the most general case, the pulse propagation
is described with a scalar equation for the transverse component
of a linearly polarized electric field. Unidirectional propagation
equations can be written in the form of generalized canonical
equations in the spectral domain for the electrical field
$\hat{E}\equiv
\hat{E}(\omega,k_x,k_y,z)$\cite{Kolesik2004a,Couairon2011}:

\begin{equation}
\frac{\partial \hat{E}}{\partial z}= iK(\omega,k_x,k_y) \hat{E}+
iQ(\omega,k_x,k_y) \frac{\hat{P}_{\mathrm{NL}}}{\epsilon_0},
\label{eq:can2}
\end{equation}

where $K(\omega,k_x,k_y) \equiv \sqrt{\omega^2 n^2(\omega)/c^2
-k_x^2 -k_y^2}$ represents the propagation constant for the modal
components of the electric field, $n(\omega)$ denotes the linear
refractive index of the medium, $c$ is the speed of light in a
vacuum, $\epsilon_0$ is the permittivity of free space and
$Q(\omega,k_x,k_y)\equiv \omega^2/2c^2K(\omega,k_x,k_y)$. The
nonlinear polarization ${P}_{\mathrm{NL}}(t,x,y,z)$, whose
spectral representation reads as
$\hat{P}_{\mathrm{NL}}(\omega,k_x,k_y,z)$, describes the nonlinear
response of the material and takes the form of constitutive
relations linking ${P}_{\mathrm{NL}}(t,x,y,z)$ to the electric
field $E(t,x,y,z)$. Expressing the propagation equation in the
canonical form has the main advantage of generality, as
Eq.~(\ref{eq:can2}) encompasses all unidirectional scalar
propagation models that can be derived under various
approximations, and can be simulated by means of a single
algorithm \cite{Couairon2011}. $E(t,x,y,z)$ represents the
electric field in the case of a carrier-resolving propagation
equation, or the envelope of the laser pulse if the slowly varying
envelope approximation is valid, i.e. if the pulse duration is
much longer than the optical cycle. Note that vectorial
unidirectional pulse propagation can also be modelled by
Eq.~(\ref{eq:can2}), for instance in tight focusing conditions,
provided $\hat{E}(\omega,k_x,k_y,z)$ is replaced by an appropriate
component of the Hertz potential \cite{Couairon2015}.

The functional forms of $K$ and $Q$ are quite general even if
slightly different forms may be in use to reflect various
approximations. For example, the paraxial approximation amounts to
performing a small $(k_x^2+k_y^2)$-expansion of $K$ and $Q$ as
$K\sim k(\omega) -(k_x^2+k_y^2)/2 k(\omega)$ and $Q\sim
\omega/2cn(\omega)$, where $k(\omega)\equiv n(\omega) \omega/c$.
In this case, Eq.~(\ref{eq:can2}) is the spectral representation
of the forward Maxwell equation \cite{Husakou2001}. For an
envelope propagation equation, $k(\omega)$ is usually expanded as
a Taylor series around the pulse carrier frequency $\omega_0$, as

\begin{equation}
k(\omega)\sim k_0+ k_0'(\omega-\omega_0)+
\frac{k_0''}{2}(\omega-\omega_0)^2+\frac{k_0'''}{3!}(\omega-\omega_0)^3
+\cdots,
\end{equation}

Combination of the paraxial approximation with the slowly varying
envelope approximation, truncation the Taylor series to the second
order (and using $K\sim k(\omega) -(k_x^2+k_y^2)/2k_0$, $Q\sim
\omega_0/2cn_0$) and factorization of the carrier wave
$\exp(-i\omega_0 t)$, leads to the family of nonlinear
Schr{\"o}dinger propagation equations:

\begin{equation}
\frac{\partial {E}}{\partial z}+ k_0'\frac{\partial {E}}{\partial
t}=\frac{i}{2k_0}\Delta_{\perp} {E}-i\frac{k_0''}{2}
\frac{\partial^2 {E}}{\partial t^2} + i\frac{\omega_0}{2 c n_0}
\frac{{P}_{\mathrm{NL}}}{\epsilon_0}. \label{eq:NLS}
\end{equation}

Including higher order terms in the Taylor series and keeping the
first order in $\omega-\omega_0$ corrective terms in $k(\omega)$
and $Q(\omega)$, the family of nonlinear envelope equations is
obtained \cite{Brabec1997}.

The term $k_0'\partial E/\partial t$ on the left hand side of
Eq.~(\ref{eq:NLS}) represents unidirectional propagation of the
pulse at the group velocity $k_0'^{-1}$. The first term on the
right hand side accounts for diffraction, where $\Delta_{\perp}$
denotes the Laplacian with respect to the coordinates $x$ and $y$
in the transverse plane. The second term accounts for the group
velocity dispersion. The sign of the coefficient $k_0''\equiv d^2
k / d \omega^2|_{\omega_0}$ determines whether dispersion is
normal ($k_0''>0$) or anomalous ($k_0''<0$). Note that its sign
also determines the hyperbolic or elliptic nature of space and
time couplings in Eq.~(\ref{eq:NLS}), which are responsible for
spatiotempral reshaping into nonlinear X- and O-waves,
respectively, as discussed in the previous section.

The nonlinear polarization is expressed in a form that
distinguishes the responses of bound and free electrons:

\begin{equation}
{P}_{\mathrm{NL}}=P_{\mathrm{bound}}+ P_{\mathrm{free}}.
\end{equation}

The response of bound electrons can be modelled via instantaneous
Kerr or delayed Raman responses, whose fractional contributions
are denoted by $\alpha$ and $1-\alpha$, respectively, assuming
that the total cubic susceptibility is constant within the
frequency range of interest:

\begin{eqnarray}
P_{\mathrm{bound}}= \epsilon_0 \chi^{(3)}\left ( \int_{-\infty}^t R(t-t')E^2(t') dt'\right ) E(t),\\
R(t)=(1-\alpha) \delta (t) + \alpha H(t) \Omega\exp(-\Gamma t)\sin
( \Lambda t),
\end{eqnarray}

where $\delta(t)$ is the Dirac delta-function, $H(t)$ is the
Heavyside step-function, and $\Omega=
\frac{\Lambda^2+\Gamma^2}{\Lambda}$, with $\Gamma$ and $\Lambda$
being the characteristic frequencies for the Raman response of the
dielectric medium. The nonlinear polarization induced by bound
electrons, $P_{\mathrm{bound}}$, is responsible for two of the
most relevant effects in filamentation and supercontinuum
generation: self-focusing and self-phase modulation (discussed in
the previous section). For a carrier resolving model, this term
also accounts for third harmonic generation and generation of
other low order odd harmonics by cascaded four wave mixing. For an
envelope propagation model designed to simulate the supercontinuum
generation over a limited spectral region, if no spectral overlap
with the third harmonic is expected, it is sufficient to replace
the field squared $E^2(t)$ by the squared modulus of the complex
envelope $|E|^2$ in the nonlinear response.

Free electron generation can be modelled by considering a simple
rate equation describing the evolution of the plasma density
$\rho_e$:

\begin{equation}
\partial_t \rho_e = W(I)(\rho_{\mathrm{nt}} -\rho_e) + \frac{\sigma}{U_g} \rho_e
I + \partial_t \rho_e|_{\mathrm{rec}},
\label{eq:rho}\end{equation}

where $W(I)$ denotes the intensity dependent photo-ionization
rate, $\rho_{\mathrm{nt}}$ is the neutral density in the valence
band, ${U_g}$ is the energy gap between the valence and the
conduction band and $\sigma$ is the cross section for inverse
Bremsstrahlung. The first term on the right hand side of
Eq.~(\ref{eq:rho}) stands for photo-ionization, the second term
stands for avalanche ionization, while the third term stands for
recombination. The response of free electrons is conveniently
described by a current, acting as a source term in the propagation
equation. The total current is linked to the nonlinear
polarization induced by free electrons, $P_{\mathrm{free}}$, and
is contributed by two components:

\begin{eqnarray}
\partial_t P_{\mathrm{free}}= J_e+ J_{\mathrm{loss}},\\
\partial_t J_e + \nu_c {J_e}= \frac{e^2}{m}\rho_e E, \label{eq:Drude}\\
J_{\mathrm{loss}}=\epsilon_0 c n_0 W(I) U_g  (\rho_{\mathrm{nt}}
-\rho_e)E. \label{eq:NLL}
\end{eqnarray}

Equation (\ref{eq:Drude}) is based on the Drude model and
describes the motion of electrons accelerated by the laser field,
undergoing friction at a rate $\nu_c$ due to collisions with ions.
$J_{\mathrm{loss}}$ is responsible for the loss of energy
necessary to ionize the medium and is described by the
phenomenological equation Eq.~(\ref{eq:NLL}). As a source term in
the propagation Eq.~(\ref{eq:can2}), $J_e$ is responsible for
plasma defocusing and plasma absorption.

For further details, an interested reader may refer to a
didactically excellent Review \cite{Couairon2011}, which provides
the necessary theoretical background, basic building blocks and
tools to perform numerical simulation with proper understanding of
the underlying physical effects. A more recent Review
\cite{Kolesik2014} provides a classification of various approaches
to optical-field evolution equations, light–matter interaction
models and methods that can be integrated with time- and
space-resolved simulations encompassing a wide range of realistic
experimental scenarios.

\section{Practical considerations}

Although femtosecond filamentation and SC generation emerge from a
complex interplay among linear (diffraction and GVD) and nonlinear
effects (self-focusing, self-phase modulation, pulse splitting,
pulse-front steepening, generation of optical shocks, multiphoton
absorption and free electron plasma generation), the practical
setup for supercontinuum generation is amazingly simple. It
involves just a focusing lens, a piece of suitable nonlinear
material and a collimating lens. In general, the laser wavelength
and the value of the nonlinear index of refraction define the
critical power for self-focusing, see Eq.~(\ref{eq:pcr}), which in
turn sets the lowest margin of the input energy required to
generate a light filament. In the near infrared spectral range,
the typical values of $P_{\rm cr}$ in wide bandgap dielectric
media are of the order of several MW, that are easily achieved
with femtosecond pulses of energies in the range from few
microjoules to few-hundreds of nanojoules. Typically, the incident
beam is externally focused to a diameter of $30-100~\mu$m, which
guarantees the location of the nonlinear focus inside the
nonlinear medium of several mm thickness. Under suitable focusing
conditions, the SC is excited with the input beam powers which
exceed $P_{\rm cr}$ just by several tens of percents. Slightly
converging or diverging laser beams may be also in use; then the
position of the nonlinear focus is defined by

\begin{equation}
\frac{1}{z'_{\rm sf}}=\frac{1}{z_{\rm sf}}+\frac{1}{f},
\end{equation}

where $f$ is the focal length of the focusing lens. Notice that
with converging input beam the nonlinear focus occurs before the
geometrical focus. In the case of diverging input beam (the
geometrical focus is located before the input face of the
nonlinear medium), the input pulse power should exceed $P_{\rm
cr}$ by several times, since the self-focusing effect should
overcome the divergence of the beam.

\begin{table*}
\caption{Linear and nonlinear parameters of basic dielectric media
used for supercontinuum generation. $U_g$ is the energy bandgap,
the transmission range is defined at $10\%$ transmission level in
1 mm thick sample, $n_0$ and $n_2$ are linear and nonlinear
refractive indexes, respectively, and are given for $\lambda=800$
nm, $\lambda_0$ is the zero GVD wavelength.}

\begin{tabular}{cccccc}
        \hline
Material&$U_g$& Transmittance & $n_2$ & $n_0$ & $\lambda_0$\\
        & eV & $\mu$m & $\times$10$^{-16}$~cm$^2$/W &  & $\mu$m\\
        \hline
        \\
LiF & 13.6\cite{Weber2003}  &$0.12-6.6$\cite{Weber2003} & 0.81\cite{Desalvo1996} & 1.39\cite{Li1976} & 1.23\cite{Li1976}\\

CaF$_2$ & 10\cite{Weber2003} &$0.12-10$\cite{Weber2003} & 1.3\cite{Adair1989} &  1.43\cite{Li1980} & 1.55\cite{Li1980}\\

Al$_2$O$_3$ & 9.9\cite{Weber2003}&$0.19-5.2$\cite{Weber2003} &3.1\cite{Major2004}& 1.76\cite{Weber2003} & 1.31\cite{Weber2003}\\

BaF$_2$ & 9.1\cite{Weber2003}&$0.14-13$\cite{Weber2003} & 1.91\cite{Adair1989} & 1.47\cite{Li1980}  & 1.93\cite{Li1980}\\

SiO$_2$ (FS) & 9.0\cite{Couairon2005}&$0.18-3.5$\cite{Weber2003} & 2.4\cite{Milam1998} & 1.45\cite{Malitson1965} & 1.27\cite{Malitson1965}\\

KDP & 7.0\cite{Nikogosyan}& $0.18-1.55$\cite{Nikogosyan} & 2.0\cite{Nikogosyan} & 1.50\cite{Zernike}& 0.98\cite{Zernike}\\

H$_2$O  & 6.9\cite{Coe2007}   & $0.18-1.3$\cite{Weber2003} & 5.7\cite{Nibbering1995} & 1.33\cite{VanEngen1998} & 1.0\cite{VanEngen1998}\\

YAG & 6.5\cite{Xu1999} &$0.21-5.2$\cite{Weber2003} & 6.2\cite{Adair1989} & 1.82\cite{Zelmon1998} & 1.60\cite{Zelmon1998}\\

$\beta$-BBO & 6.2\cite{Nikogosyan}&$0.19-3.5$\cite{Nikogosyan} & 5.2\cite{Bache2013} & 1.66\cite{Zhang2000} & 1.49\cite{Zhang2000}\\

BK7 & 4.28\cite{Little2011} & $0.3-2.5$\cite{SCHOTT} & 3.75\cite{Nibbering1995} & 1.51\cite{SCHOTT} & 1.32\cite{SCHOTT}\\

KGW & 4.05\cite{Major2003} & $0.3-5$\cite{Webb2004} & 11\cite{selivanov2006} & 2.02\cite{Pujol1999} & 2.2\cite{Pujol1999}\\

YVO$_4$ & 3.8\cite{Dolgos2009} &$0.35-4.8$\cite{Weber2003} & 15\cite{selivanov2006} & 1.97\cite{Weber2003} & \\

        \hline
    \end{tabular}
\end{table*}

\begin{figure}[ht!]
\includegraphics[width=8.5cm]{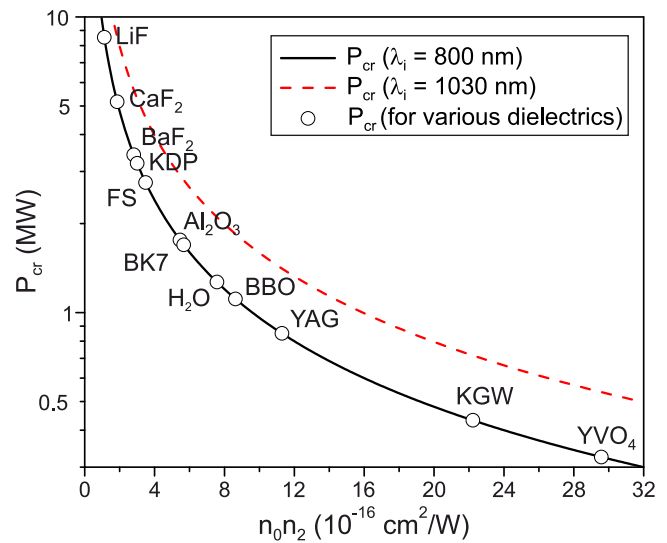}
\caption{Critical power for self-focusing calculated for
Ti:sapphire (800 nm, solid curve) and Yb-doped (1030 nm, dashed
curve) laser wavelengths. Circles denote the critical power for
self-focusing at 800 nm in various dielectric
materials.}\label{fig:pcr}
\end{figure}

Relevant linear and nonlinear parameters for widely used nonlinear
dielectric media are provided in Table~1. Figure~\ref{fig:pcr}
illustrates the calculated critical power for self-focusing in
various dielectric materials at the input wavelengths of 800 nm
and 1030 nm, which are emitted by commonly used Ti:sapphire and
Yb-doped femtosecond lasers, respectively.

\begin{figure}[ht!]
\includegraphics[width=8.5cm]{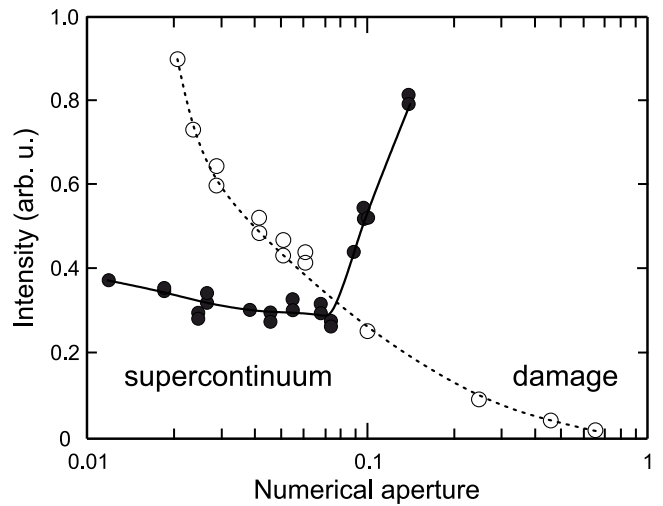}
\caption{Energy thresholds for damage (open circles) and
supercontinuum generation (filled circles) in UV-grade fused
silica with 60 fs, 800 nm laser pulses versus NA. The curves serve
as guides for the eye. Adapted from
\cite{Ashcom2006}.}\label{fig:damage}
\end{figure}

However, to generate stable and reproducible SC, some important
issues of laser-matter interaction, and femtosecond filamentation
in particular, should be taken into consideration. These practical
issues are universal and hold for any nonlinear medium and at any
input wavelength in the optical range. First of all, the optical
damage of the medium is the major limiting factor in SC
generation, therefore the external focusing condition (the
numerical aperture, NA) should be carefully chosen
\cite{Nguyen2003,Ashcom2006}. Figure~\ref{fig:damage} shows the
experimentally measured threshold energies for SC generation and
optical damage in fused silica as functions of the numerical
aperture \cite{Ashcom2006}.

In the high-NA regime (NA$>0.25$), the optical damage occurs for
the input pulse energies below the energy corresponding to the
critical power of self-focusing, so no SC generation under such
focusing condition is observed. Most of the pulse energy is thus
deposited into the material at the focal volume through
multiphoton absorption and subsequent linear absorption by the
plasma; a laser-matter interaction regime that is exploited for
micromachining of transparent bulk materials. In the NA range from
0.15 to 0.05, the threshold energies for SC generation and optical
damage are very close. In particular, for NA$<0.1$, the SC is
generated without the optical damage in a single shot regime,
however the damage accumulates under multiple shot exposure,
causing the SC to disappear over time. Finally, with NA below
0.05, it is still possible to damage the medium, but only at the
input pulse energies significantly above the threshold for SC
generation, hence constituting the ``safe'' operating condition
for SC generation. Moreover, experiments show that loose focusing
condition facilitates enhanced red-shifted broadening of the SC
spectrum \cite{Bradler2009}, which stems from increased nonlinear
propagation of the leading sub-pulse, which preserves a steep
ascending front \cite{Jukna2014}.

\begin{figure}[ht!]
\includegraphics[width=8.5cm]{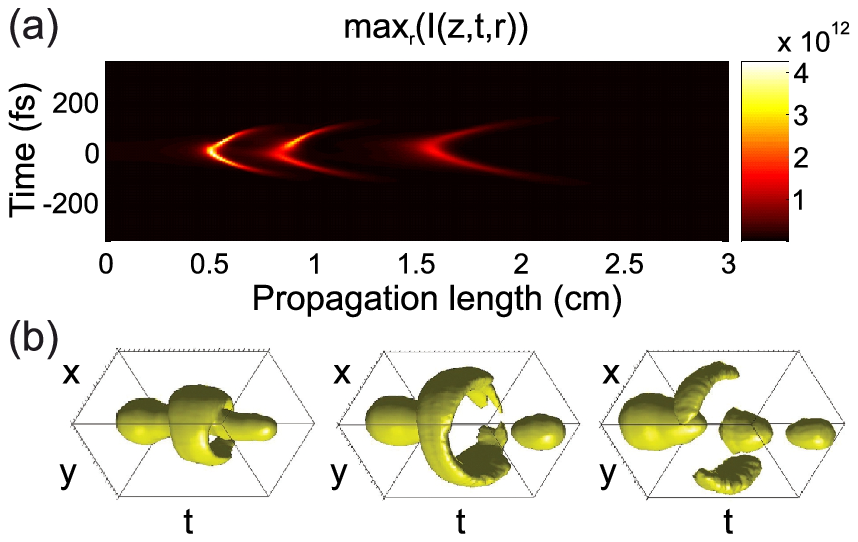}
\caption{(a) Numerical simulation of temporal profiles for the
axial intensity distribution illustrating focusing/refocusing
cycles and recurrent pulse splitting events of 90 fs ultraviolet
(400 nm) laser pulse with 400 nJ energy propagating in water. (b)
Experimentally measured spatiotemporal $(x,y,t)$ profiles of a
filament at various stages of propagation after the first
nonlinear focus. Adapted from
\cite{Jarnac2014}.}\label{fig:focdefoc}
\end{figure}

However, for numerical apertures below 0.05, that correspond to
loose focusing condition, sufficiently long nonlinear medium and
the input beam power exceeding $P_{\rm cr}$ by several times,
femtosecond filament may undergo recurrent self-focusing cycles,
which could be directly visualized by monitoring the multiphoton
absorption-induced fluorescence traces with specific color
appearance, which depends on the material, as detected from a side
view of the nonlinear medium, see e.g.
\cite{Wu2002,Liu2003b,Dharmadhikari2009}. The general
interpretation of focusing/refocusing cycles is based on the
so-called dynamic spatial replenishment scenario, which assumes
alternating cycles of self-focusing due to Kerr effect and
self-defocusing due to free electron plasma and which was
originally proposed to explain the illusion of long-distance
self-guided propagation of high-power pulses in gaseous media
\cite{Mlejnek1998}. More recent experimental and numerical study
of full spatiotemporal evolution of light filaments versus the
propagation distance in water unveiled the intimate connections
between complex propagation effects: focusing and refocusing
cycles, nonlinear absorption, pulse splitting and replenishment,
supercontinuum generation and conical emission \cite{Jarnac2014}.
More specifically, whenever the self-focusing wave packet (the
ultrashort pulsed laser beam) approaches the nonlinear focus,
multiphoton absorption attenuates its central part, which after
the pulse splitting is reshaped into a ring-like structure. With
further propagation the leading and trailing sub-pulses separate
before the light contained in the ring refocuses and replenishes
the pulse on the propagation axis, as illustrated in
Fig.~\ref{fig:focdefoc}(b). If the power of the replenished pulse
is above critical, the replenished pulse undergoes another
self-focusing cycle, which results in pulse splitting at the
secondary nonlinear focus. The secondary pulse splitting produces
yet another portion of the SC, and the resulting SC spectrum and
conical emission pattern develop a periodic modulation, as due to
interference between the primary and the secondary split
sub-pulses. After the tertiary splitting event, the modulation in
the SC spectrum has beatings contributed by the occurrence of a
tertiary split sub-pulses, etc. Focusing/refocusing cycles may
repeat as long as the replenished pulse contains the power still
above critical. However, each refocusing cycle is followed by a
sudden decrease in the transmittance and therefore the entire beam
will continuously lose energy during propagation, and eventually a
linear propagation regime is resumed.

\begin{figure}[ht!]
\includegraphics[width=8.5cm]{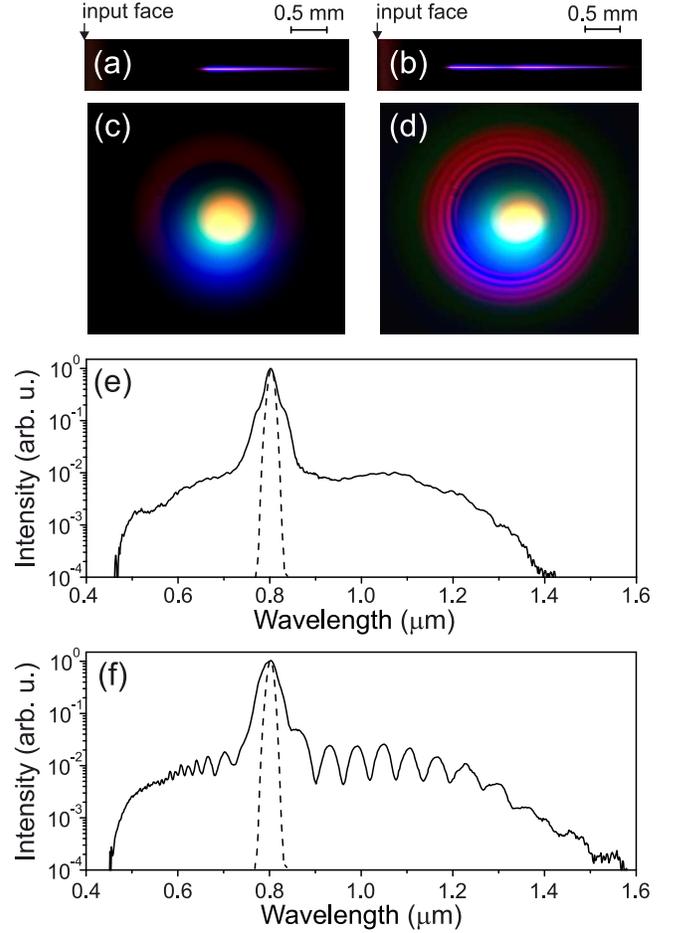}
\caption{Plasma fluorescence traces in a YAG crystal induced by
self-focusing of 100 fs, 800 nm input pulses with energies of (a)
310 nJ, (b) 560 nJ, that induce a single self-focusing event and
refocusing of the filament, respectively. (c) and (d) show the
corresponding far-field patterns of SC emission, (e) and (f) show
the corresponding SC spectra. Dashed curves show the input pulse
spectrum. See text for details.}\label{fig:yag800}
\end{figure}

Figure~\ref{fig:yag800} illustrates the above considerations by
comparing visually perceptible filamentation features and SC
spectra in a YAG crystal, as generated with 100 fs, 800 nm input
pulses with energies of 310 nJ (peak power of 3.6 $P_{\rm cr}$)
and 560 nJ (6.6 $P_{\rm cr}$), that induce a single self-focusing
event and refocusing of the filament, respectively. The single
self-focusing event is visualized by a gradually decaying plasma
fluorescence trace, whose most intense part indicates the position
of the nonlinear focus, as shown Fig.~\ref{fig:yag800}(a), and
produces featureless far-field pattern of SC emission
[Fig.~\ref{fig:yag800}(c)] and smooth SC spectrum
[Fig.~\ref{fig:yag800}(e)]. In contrast, the refocusing of the
filament emerges as a double-peaked plasma fluorescence trace, as
shown in Fig.~\ref{fig:yag800}(b), and results in the occurrence
of modulation in the outer part of the far-field pattern of SC
emission [Fig.~\ref{fig:yag800}(d)] and periodic modulation of the
SC spectrum [Fig.~\ref{fig:yag800}(f)]. These indications are very
important from a practical point of view, since they allow us to
easily identify the recurrent collapse and pulse splitting events
and optimize the operating conditions for SC generation without
employing complex experimental measurements.

\begin{figure*}[ht!]\center
\includegraphics[width=17cm]{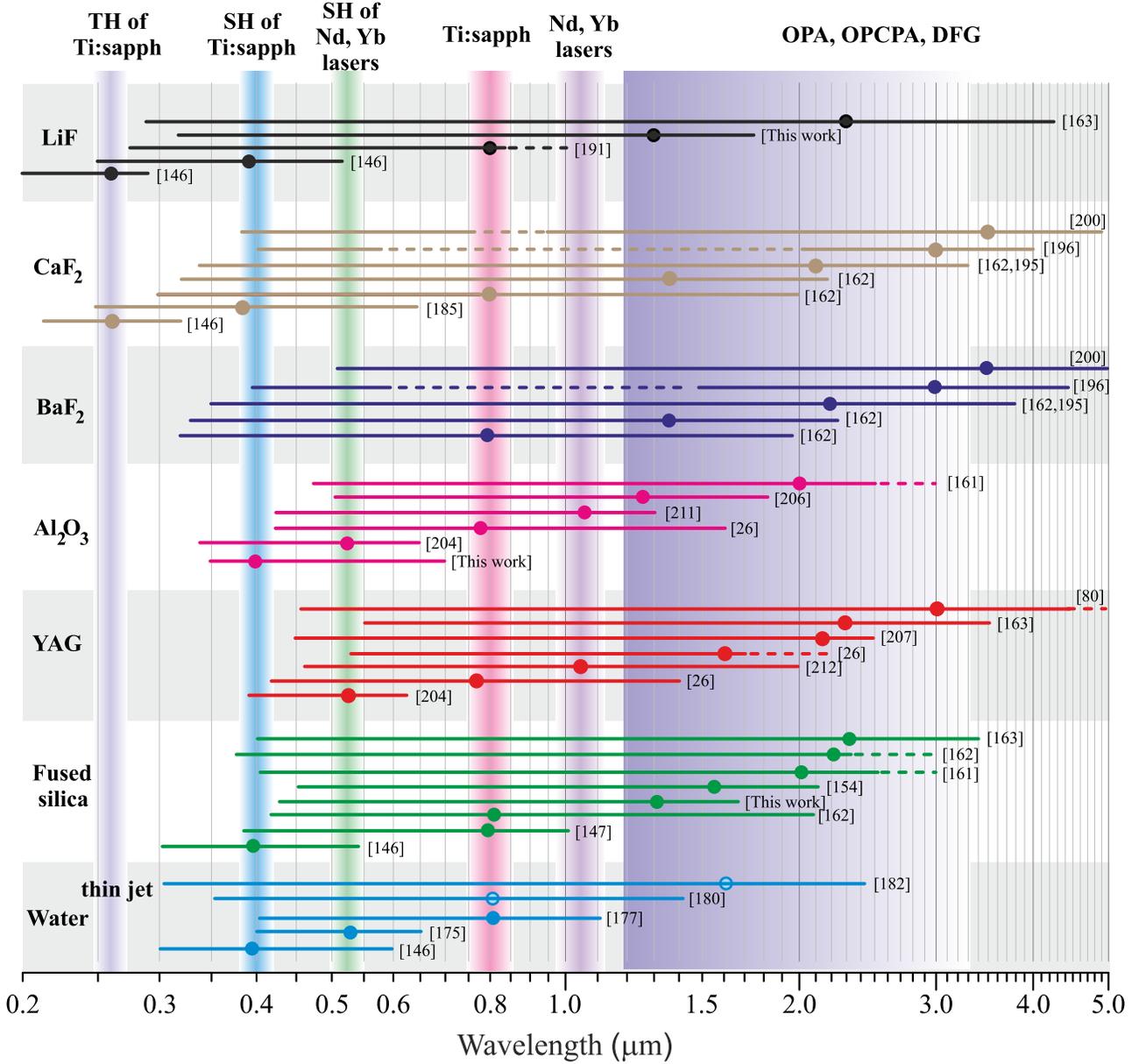}
\caption{Summary of the experimentally measured supercontinuum
spectra in commonly used wide bandgap dielectric media. The bold
circles mark the pump wavelengths. Dashed lines indicate the
spectral ranges into which further spectral broadening was
expected, but not measured because of limited detection range.
Vertical color shading and labels on top indicate the laser
sources: TH, third harmonic; SH, second harmonic; OPA, optical
parametric amplifier; OPCPA, optical parametric chirped pulse
amplifier; DFG, difference frequency
generator.}\label{fig:summary}
\end{figure*}

In the filamentation regime in the range of anomalous GVD, where
pulse self-compression rather than splitting at the nonlinear
focus takes place, the recurrent self-focusing cycles manifest
themselves in a similar manner \cite{Chekalin2015,Kuznetsov2016}.
The refocusing event was shown to produce splitting of the light
bullet at the secondary nonlinear focus \cite{Grazuleviciute2015},
while multiple refocusing cycles were demonstrated to yield
formation of a sequence of quasi-periodical light bullets, each of
them resulting in ejection of the new portion of the SC.

Finally, further increase of the input pulse energy leads to beam
break-up into several or multiple filaments, which emerge from
random amplitude and phase fluctuations inevitably present in real
input beams and pulses. The intensity distributions of multiple
filaments emerge in the form of regular or irregular patterns,
that are governed by the input beam size, symmetry and smoothness,
see e.g. \cite{Dubietis2004,Majus2010,Berge2010}. Although the
multifilamentation regime creates an illusion of energy scaling of
SC, each individual filament is subject to a combined effect of
the intensity clamping \cite{Liu2002a} and the dispersion
landscape of the material \cite{Kolesik2003a,Kolesik2003b}, hence
no additional spectral broadening with increasing the input power
is achieved. Moreover, beam break-up into multiple filaments
deteriorates the spatial uniformity of the beam and the temporal
structure of the pulse, and eventually induces a considerable
depolarization at various parts of the SC spectrum
\cite{Dharmadhikari2006c,Kumar2008}.

\section{Supercontinuum generation in wide bandgap dielectrics}

Figure~\ref{fig:summary} presents a graphical summary of the most
relevant experimental results on SC generation in wide bandgap
dielectric media, as achieved with various femtosecond laser
sources spanning pump wavelengths from the ultraviolet to the
mid-infrared. The experimental details are provided below.

\subsection{Glasses}

Silica-based glasses are multifunctional optical materials, which
find diverse applications in contemporary optical sciences and
technology. Thanks to a large bandgap and reasonably large
nonlinear refractive index, as combined with high optical and
mechanical quality, fused silica serves as an etalon nonlinear
medium for studies of various fundamental and practical aspects of
laser-matter interaction, such as self-focusing, self-phase
modulation, photoionization, free-carrier absorption,
carrier-carrier interaction, and eventually, mechanisms of the
optical damage \cite{Schaffer2001,Tzortzakis2001,Couairon2005}.
Therefore it is of no surprise that experiments in fused silica
provided a valued practical knowledge, which allowed optimization
of experimental schemes for SC generation.

Spectral broadening and SC generation in fused silica was
investigated within a wide range of pump wavelengths, from the
deep ultraviolet to the mid-infrared. In the deep ultraviolet,
only very slight spectral broadening around the carrier wavelength
was observed with femtosecond pulses at 248 nm from an amplified
excimer laser \cite{Tzortzakis2006} and with pulses at 262 nm
produced by frequency tripling of the Ti:sapphire laser output
\cite{Nagura2002}. More noticeable, but still modest spectral
broadening in fused silica and BK7 glass was reported with the
second harmonic pulses at 393 nm of the Ti:sapphire laser
\cite{Nagura2002}. With the pump wavelength located in the visible
spectral range, the SC spectrum from 415 nm to 720 nm was
generated with 250 fs pulses at 594 nm from a rhodamine 6G dye
laser \cite{Wittmann1996}.

\begin{figure}[ht!]
\includegraphics[width=8.5cm]{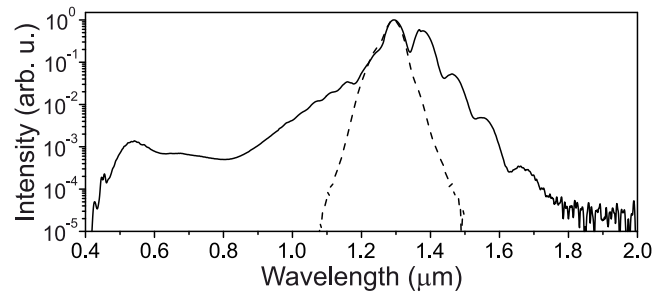}
\caption{Supercontinuum spectrum generated by filamentation of 100
fs pulses with an energy of $1.9~\mu$J in 3 mm-long UV-grade fused
silica sample. The dashed curve shows the input pulse spectrum,
whose central wavelength (1.3 $\mu$m) is close to the zero GVD
point of fused silica.}\label{fig:fs13}
\end{figure}

Many experiments on SC generation in fused silica were performed
with a standard near-infrared pumping, using the fundamental
harmonic of the Ti:sapphire laser and with the input pulsewidths
of 100 fs and shorter. A typical SC spectrum produced in fused
silica extends from 390 nm to 1000 nm, as reported under various
experimental conditions
\cite{Nagura2002,Nguyen2003,Fang2003,Ashcom2006,Dachraoui2010,Zhang2016}.
Although similar, but somewhat narrower SC spectra were generated
in BK7 \cite{Nagura2002,Dharmadhikari2005,Dharmadhikari2006b} and
ZK7 \cite{Yang2007} glasses, which possess smaller bandgaps.
Filamentation of laser pulses with longer wavelengths, as
delivered by the optical parametric amplifiers, yielded
considerably broader SC spectra, as reported with pump wavelengths
close or slightly above the zero GVD points of fused silica and
BK7 glass \cite{Saliminia2005,Naudeau2006,Jiang2015}, which are
located at 1.27 $\mu$m and 1.32 $\mu$m, respectively, see Table~1.
Figure~\ref{fig:fs13} shows an example of the SC spectrum
generated at near zero GVD point of fused silica (the pump
wavelength 1.3 $\mu$m).

\begin{figure}[ht!]
\includegraphics[width=8.5cm]{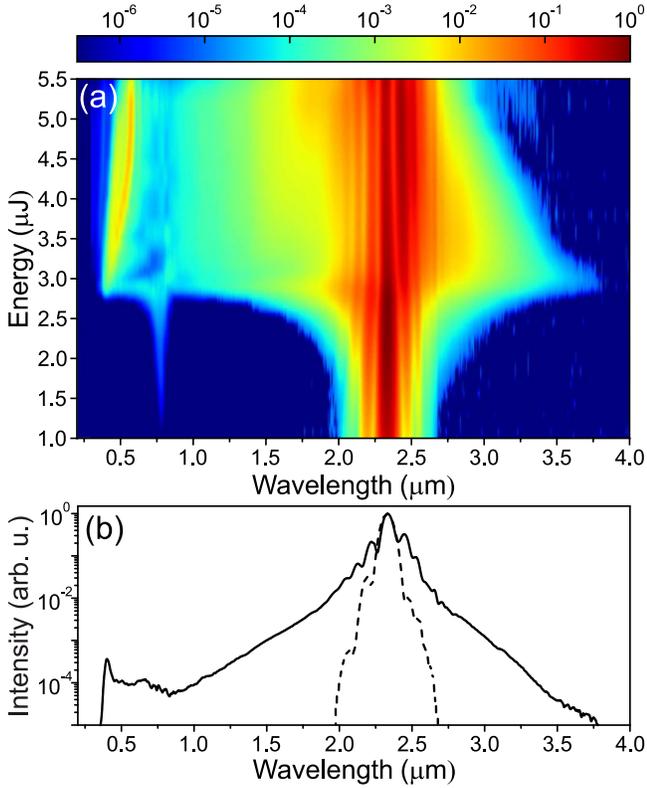}
\caption{(a) Spectral broadening of 100 fs, 2.3 $\mu$m laser pulse
in 3-mm-long fused silica sample versus the input pulse energy.
The spectral peak centered at 767 nm, which appears before the
onset of supercontinuum generation, is the third harmonic. (b) The
broadest supercontinuum spectrum as measured with the input pulse
energy of 2.8 $\mu$J. Notice an intense blue peak centered at 430
nm. The input pulse spectrum is shown by the dashed curve. Adapted
from \cite{Garejev2017}.}\label{fig:fs}
\end{figure}

SC generation experiments with pump wavelengths falling in the
range of anomalous GVD of fused silica uncovered a number of
universal features, which characterize the entire shape of the SC
spectrum in the spectral-angular domain
\cite{Faccio2006,Porras2007,Durand2013b}. The angle-integrated as
well as the axial SC spectra show an intense blue-shifted peak
located in the visible range, which is identified as an axial
component of the conical emission and whose blue-shift increases
with increasing the wavelength of the driving pulse
\cite{Smetanina2013,Durand2013b}. Various aspects of the spectral
broadening and SC generation, such as energy content, stability of
the carrier envelope phase, etc., were studied in connection to
formation and propagation dynamics of spatiotemporal light bullets
\cite{Chekalin2013,Chekalin2015a,Grazuleviciute2015a}. Experiments
on self-focusing and filamentation in fused silica of few optical
cycle pulses with carrier wavelengths of 2 $\mu$m
\cite{Darginavicius2013} and 2.2 $\mu$m \cite{Dharmadhikari2014}
reported ultrabroad SC spectra starting from 400 nm and 370 nm,
respectively, and extending to wavelengths greater than 2.5
$\mu$m. More recently, the recorded dynamics of spectral
broadening in fused silica with the pump wavelength of 2.3 $\mu$m
versus the input pulse energy revealed that under given
experimental conditions, there exists an optimum pump pulse energy
to obtain the broadest SC spectrum \cite{Garejev2017}, see
Fig.~\ref{fig:fs}(a). Figure~\ref{fig:fs}(b) shows the broadest SC
spectrum generated with the optimum input pulse energy of 2.8
$\mu$J, providing a continuous wavelength coverage from 310 nm to
3.75 $\mu$m, which converts to 3.6 optical octaves.

Various non-silica glasses currently receive an increasing
attention as potential candidates for SC generation. Using the
pump pulses at 1.6 $\mu$m from the optical parametric amplifier,
the SC spanning wavelengths from 400 to 2800 nm was reported in
lanthanum glass \cite{Yang2017}. An impressive spectral broadening
was observed in fluoride glass (ZBLAN) under similar pumping
conditions; the authors generated an ultra-broadband, more than 5
octaves-wide SC, covering almost the entire transmission range of
the material ($0.2-8.0~\mu$m) \cite{Liao2013a}. The so-called soft
glasses hold a great potential for SC generation in the mid
infrared spectral range. These materials are widely exploited in
modern optical fiber technology, however, bulk soft glasses, such
as tellurite and chalcogenide, also show very promising results.
The SC spectra extending from the visible to 6 $\mu$m and from the
visible to 4 $\mu$m were reported in tellurite glass using the
pump pulses with central wavelengths of 1.6 $\mu$m
\cite{Liao2013b} and 2.05 $\mu$m \cite{Bejot2016}, respectively. A
flat SC in the $2.5-7.5~\mu$m range was produced in bulk
chalcogenide glass using the pump pulse with central wavelength of
5.3 $\mu$m \cite{Yu2013}. The SC spectra with remarkable
mid-infrared coverage from 2.5 to $\sim11$ $\mu$m were produced in
As$_2$S$_3$ and GeS$_3$ bulk chalcogenide glass samples using 65
fs pulses at 4.8 $\mu$m \cite{Mouawad2015}. More recent study
reported on generation of the mid-infrared SC extending from 2.44
to 12 $\mu$m in chalcogenide glass samples of various composition,
as pumped with 70 fs pulses of wavelength tunable in the
$3.75-5.0~\mu$m range \cite{Stingel2017}.

\subsection{Water}

The absence of permanent optical damage and the possibility to
easily vary the medium thickness regards liquids as attractive
media for many experiments in nonlinear optics. Water occupies an
exceptional place among the other liquids because of its
technological and biomedical importance. Alongside fused silica,
in many occasions water serves as a prototypical nonlinear medium
for studies of ultrafast light-matter interactions. Systematic
studies of the spectral broadening in water date back to the mid
1980s \cite{Smith1977} and successfully continue in the
femtosecond laser era, see
\cite{Golub1990,He1993,Brodeur1996,Wittmann1996} for the early
accounts on femtosecond SC generation in water and some other
liquids.

Generation of femtosecond SC in water was performed with
ultraviolet, visible and near infrared pump wavelengths. With the
ultraviolet pump pulses generated by the frequency doubling of an
amplified Ti:sapphire laser output, the measured SC spectra
covered wavelength ranges of $290-530$ nm  \cite{Nagura2002} and
$350-550$ nm \cite{Jarnac2014}, as reported with the pump
wavelengths of 393 nm and 400 nm, respectively, and under slightly
different external focusing conditions. With the pump wavelengths
in the visible spectral range, the SC spectrum from 400 nm to 650
nm was generated with 527 nm self-compressed femtosecond second
harmonic pulses from the Nd:glass laser \cite{Dubietis2003} and
from 450 nm to 720 nm with 594 nm pulses from a rhodamine 6G dye
laser \cite{Wittmann1996}. A considerable effort was dedicated to
study the SC generation in water with Ti:sapphire laser pulses,
under various operating conditions
\cite{Liu2002b,Liu2003a,Kandidov2003,Cook2003,Liu2005}. These
studies demonstrated that a typical SC spectrum covers the
$400-1100$ nm range \cite{Liu2003a}. A broader SC spectrum
spanning wavelengths from 350 nm to 1400 nm was reported using a
thin water jet instead of a thick cell \cite{Tcypkin2016}.

In the range of anomalous GVD of water (with laser wavelengths
above 1 $\mu$m), the SC spectrum with focusing
condition-controllable position of the blue peaks was demonstrated
as pumped by the fundamental harmonics of the Nd:glass laser (1055
nm) \cite{Faccio2008}. However, the red shifted portion of the SC
spectrum was not measured, as the spectral detection range in
these experiments was limited to that of a conventional silicon
detector (1.1 $\mu$m). Although rapidly increasing infrared
absorption and presence of strong absorption bands at 1.46 and
1.94 $\mu$m in particular, serve as the main factors that limit
the red-shifted broadening of the SC spectrum in water, an
unexpectedly broad SC spectrum, ranging from 350 to 1750 nm was
reported using 1.3 $\mu$m pump pulses from the optical parametric
amplifier \cite{Vasa2014}. More recent study suggested that under
carefully chosen experimental conditions, the absorption effects
can be overcome using a thin water jet, thus reducing the
nonlinear interaction length just to a few tens of micrometers
\cite{Dharmadhikari2016}. Under these settings, with the pump
wavelength set at 1.6 $\mu$m, more than two-octave spanning SC
spectrum from 300 nm to 2.4 $\mu$m was measured.

\subsection{Alkali metal fluorides}

Alkali metal fluorides possess the largest bandgaps among the
dielectric materials. Consequently, these media exhibit extremely
broad transparency windows, which extend from the vacuum
ultraviolet to the mid infrared, making them attractive nonlinear
materials for SC generation as pumped with femtosecond laser
pulses at various parts of the optical spectrum. In particular,
lithium and calcium fluorides, LiF and CaF$_2$, are the only
nonlinear media, that are able to produce an appreciable spectral
broadening in the deep ultraviolet, as demonstrated using third
harmonic pulses from Ti:sapphire laser
\cite{Nagura2002,Riedle2013}. As pumped with the second harmonic
pulses from the same laser, the measured SC spectra in these media
cover the ultraviolet and visible spectral range
\cite{Nagura2002,Ziolek2004,Johnson2009}.

Extensive experimental studies of SC generation in alkali metal
fluorides were performed with fundamental harmonic pulses from the
Ti:sapphire laser, with an emphasis on the blue-shifted part of
the SC spectrum. A comparative study of the SC generation in
CaF$_2$, LiF and MgF$_2$ crystals demonstrated that LiF and
CaF$_2$ crystals produce the most ultraviolet-shifted
supercontinua with the cut-off wavelengths below 300 nm
\cite{Tzankov2002}. Out of these, CaF$_2$ was identified as a
promising material for stable SC generation in the ultraviolet and
visible spectral range, as it exhibits low induced depolarization
\cite{Buchvarov2007,Kartazaev2008,Johnson2009} and excellent
reproducibility of the spectrum, as important for femtosecond
transient absorption spectroscopy
\cite{Zeller2000,Ziolek2004,Megerle2009,Krebs2013,Riedle2013} and
for improving the performance characteristics of the UV-pumped
noncollinear optical parametric amplifiers
\cite{Tzankov2002,Huber2001}. However, CaF$_2$ has a relatively
low optical damage threshold, which compares to that of fused
silica, therefore a reliable and reproducible SC generation in
this material is achieved only in the setups, where continuous
translation or rotation of the crystal is performed. Although
neglected in some studies \cite{Dachraoui2010}, formation of
persistent color centers in LiF was generally considered as a
major drawback to its application for SC generation. However, a
more recent study revealed that color centers only slightly modify
the ultraviolet cut-off (at 270 nm) of the SC spectrum on the
long-term (several hours) operation \cite{Landgraf2013}.

SC generation in barium fluoride, BaF$_2$, with 800 nm pumping was
studied under various experimental conditions
\cite{Dharmadhikari2004,Dharmadhikari2007}. BaF$_2$ is a
well-known scintillator possessing two strong luminescence bands
centered at 330 and 200 nm. The former luminescence band was
readily employed to map the intensity variation within the light
filaments; in particular, a side-view of six-photon
absorption-induced luminescence allowed to capture the filament
formation dynamics, focusing and refocusing cycles, and eventually
to estimate a number of relevant parameters, such as filament
diameter, peak intensity, free electron density and multiphoton
absorption cross section
\cite{Dharmadhikari2006a,Dharmadhikari2009}. From a practical
viewpoint, SC generation in BaF$_2$ was used to provide a
broadband seed signal for OPCPA pumped by high repetition rate
Yb:YAG thin disk regenerative amplifier; after amplification the
pulses were compressed down to 4.6 fs, which is very close to a
Fourier limit of the amplified SC spectrum \cite{Harth2012}.

More recent study of the SC generation in CaF$_2$ and BaF$_2$
captured the entire SC spectra, demonstrating a considerable
red-shifted broadening, which extends well beyond the detection
range of standard Si detectors \cite{Dharmadhikari2014}. More
specifically, with 800 nm pumping, the SC spectra extending from
300 nm to 2 $\mu$m and from 320 nm to 1.98 $\mu$m were measured in
CaF$_2$ and BaF$_2$, respectively, and further red-shifted
spectral broadening in these nonlinear crystals was recorded using
the pump pulses at 1.38 $\mu$m.

\begin{figure}[ht!]
\includegraphics[width=8.5cm]{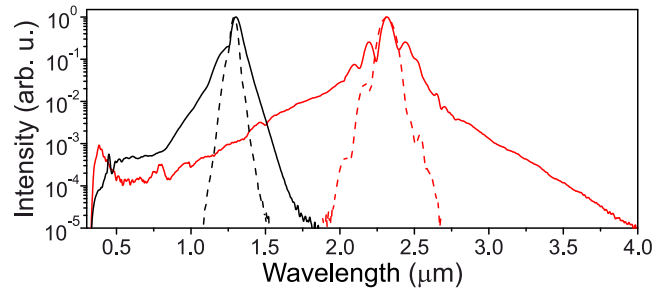}
\caption{A comparison of supercontinuum spectra in 3.5 mm-thick
LiF as produced by filamentation of 100 fs pulses with wavelengths
of $1.3~\mu$m (black curve) and $2.3~\mu$m (red curve), which fall
into the ranges of zero and anomalous GVD of the crystal and with
energies of 2.2 $\mu$J and 9.5 $\mu$J, respectively. Spectra of
the input pulses are shown by the dashed curves.}\label{fig:lif}
\end{figure}

Using the mid infrared ($2.1-2.2~\mu$m) pump pulses, whose
wavelengths fall into the range of anomalous GVD of the fluoride
crystals, combined data from \cite{Dharmadhikari2014} and
\cite{Liang2015} yield ultrabroad, multioctave SC spectra in
CaF$_2$ and BaF$_2$, which span from 340 nm to 3.3 $\mu$m and from
350 nm to 3.8 $\mu$m, respectively. The measurements performed
with 15 fs, 2 $\mu$m pump pulses revealed an exceptionally flat
shape of the SC spectrum in CaF$_2$, showing a broad plateau in
the wavelength range of $500-1700$ nm \cite{Darginavicius2013}.
Even broader, almost 4 octave-spanning SC spectrum, continuously
covering the wavelength range from 290 nm to 4.3 $\mu$m (at the
$10^{-6}$ intensity level) was reported in LiF with 2.3 $\mu$m
pump pulses \cite{Garejev2017}. Interestingly, such an
ultrabroadband spectrum was recorded in the presence of color
centers. The time-resolved evolution of the SC spectra
demonstrated that spectral modifications due to formation of color
centers evolve on a very fast time scale (just a few tens of laser
shots). However, after few thousands of laser shots the SC
spectrum eventually stabilizes and remains unchanged during
further operation. Figure~\ref{fig:lif} compares the SC spectra in
3.5 mm-long LiF plate, as generated by 100 fs pulses with
wavelengths of $1.3~\mu$m and $2.3~\mu$m, which fall into the
ranges of zero and anomalous GVD of the crystal, respectively and
showing a stable ultraviolet cut-off at 330 nm (at the $10^{-5}$
intensity level).

\begin{figure}[ht!]
\includegraphics[width=8.5cm]{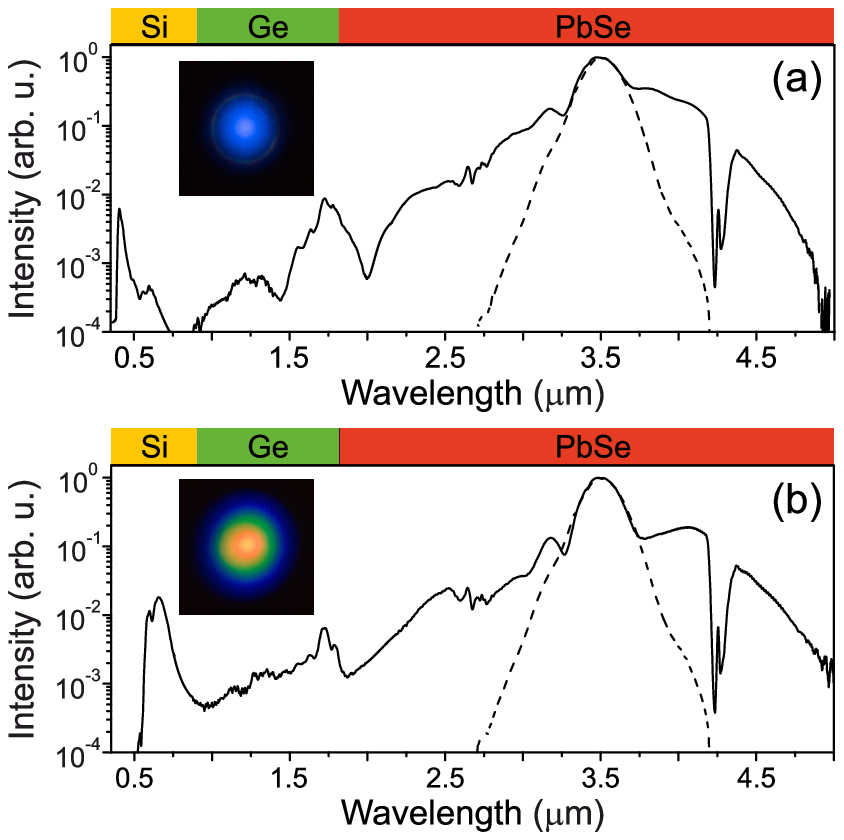}
\caption{Supercontinuum spectra generated with 60 fs, 31 $\mu$J
pulses at 3.5 $\mu$m in (a) CaF$_2$ and (b) BaF$_2$, both of 4 mm
thickness. Dashed curves show the input pulse spectrum. A deep
double dip around 4.25 $\mu$m is due to absorption of atmospheric
CO$_2$. The ranges of spectrometer detectors (Si, Ge, PbSe) are
indicated by color bars on the top. The insets show the visual
appearances of the SC beams in the far field. Adapted from
\cite{Marcinkeviciute2017}.}\label{fig:cafbaf35}
\end{figure}

The experiments on SC generation in CaF$_2$ and BaF$_2$ with
longer wavelength mid-infrared pulses (tunable in the $2500-3800$
nm range) reported a dramatic change of the SC spectral shape
\cite{Dormidonov2015,Chekalin2017a}. These measurements showed
that, for instance, with 3 $\mu$m pump pulses, the SC spectra are
no longer continuous, but are composed of two separate bands, one
located in the visible range ($400-600$ nm) and another in the mid
infrared ($1500-4500$ nm), while the spectral components in the
near infrared spectral range are practically absent or at least
their spectral intensities fall below the detection range. Similar
results were reported in LiF crystal, also demonstrating how the
spectral position of the detached blue peak (that corresponds to a
visible SC band in the above experiments) was shifting from 270 nm
to 500 nm while changing the input wavelength from 3.3 to 1.9
$\mu$m \cite{Dormidonov2016,Chekalin2017b}. More recent spectral
measurements with 3.5 $\mu$m pulses (Fig.~\ref{fig:cafbaf35}), as
performed over higher dynamic range, revealed that the spectral
discontinuity still exists in CaF$_2$, however, demonstrating more
homogenous SC spectrum in BaF$_2$, which provided the spectral
coverage from the visible to beyond $5~\mu$m in the mid-infrared
\cite{Marcinkeviciute2017}.

\subsection{Laser hosts}

Laser host crystals, such as undoped sapphire (Al$_2$O$_3$) and
yttrium aluminum garnet (YAG) are excellent nonlinear media
possessing good crystalline quality, high nonlinearity and high
optical damage threshold. Therefore it is quite surprising that
the potential of these laser host materials for SC generation was
systematically overlooked in the early studies.

The first experimental demonstration of the SC generation in
sapphire dates back to 1994 \cite{Reed1994}, revealing sapphire as
a long sought solid-state material to replace at that time
commonly used liquid media, putting the technology of femtosecond
optical parametric amplifiers on all-solid state grounds, see also
\cite{Reed1995} for more details. Since then, sapphire became a
routinely used nonlinear medium for SC generation with Ti:sapphire
driving lasers, providing high quality seed signal that boosted
the development of modern ultrafast optical parametric amplifiers
\cite{Cerullo2003,Brida2010,Manzoni2016}.

\begin{figure}[ht!]
\includegraphics[width=8.5cm]{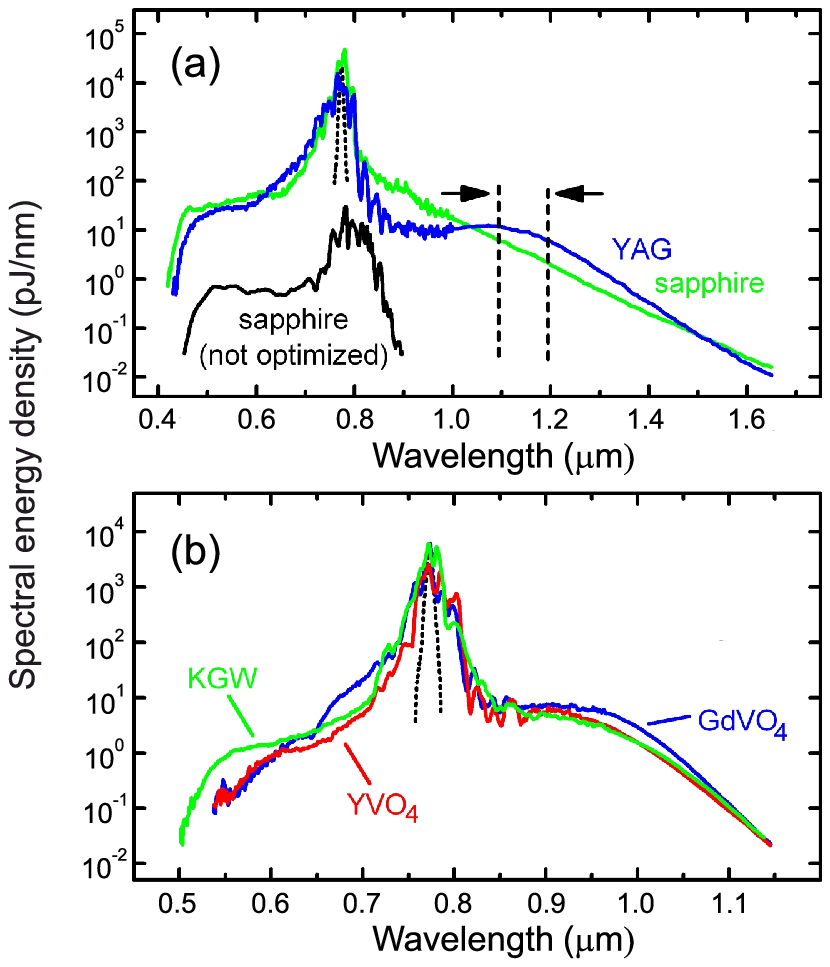}
\caption{(a) Black curve: supercontinuum spectrum from 3 mm
sapphire with conventional (tight focusing) pumping conditions
(shown not to scale, shifted for clarity). Green curve:
supercontinuum spectrum from 3 mm sapphire with loose focusing
condition. Blue curve: supercontinuum spectrum from 4 mm YAG
showing an improved photon density in the infrared region. (b)
Supercontinuum spectra generated in KGW (green curve), YVO$_4$
(red curve) and GdVO$_4$ (blue curve) crystals (all 4 mm) with the
input pulse energies of 83 nJ, 59 nJ and 78 nJ, respectively.
Adapted from \cite{Bradler2009}.}\label{fig:laserhost}
\end{figure}

Pumping sapphire crystal with the second harmonics of Ti:sapphire
laser (400 nm), the SC spectrum in the $350-700$~nm range was
produced, please refer to Fig.~\ref{fig:cofil}(a), which appears
in Section 6 of the paper. In the visible spectral range, using
the second harmonics of an amplified Yb-fiber laser (515 nm) as a
pump, the SC spectrum from 340 nm to 650 nm was measured, which
was thereafter used to seed the noncollinear optical parametric
amplifier pumped by third harmonic (343 nm) of the same laser,
rendering unprecedented tuning in the near-ultraviolet and blue
spectral ranges \cite{Bradler2014b}. A typical SC spectrum in
sapphire covers the wavelength range from 410 nm to 1100 nm, as
generated with fundamental harmonic pulses from Ti:sapphire laser
(the pump wavelengths of around 800 nm) and measured under the
commonly used operating conditions (the input pulse energy of
$\sim 1~\mu$J and the material thickness of $1-3$ mm)
\cite{Bradler2009,Majus2011,Imran2012,Majus2013}. A notable
extension of the infrared part of the SC spectrum was demonstrated
by employing looser focusing geometry and somewhat longer sapphire
samples; under these focusing conditions an appreciable
red-shifted SC signal extended to more than 1600 nm
\cite{Bradler2009,Jukna2014}, see Fig.~\ref{fig:laserhost}(a). A
broader, spectrally flat SC was produced using pump wavelengths
around the zero GVD point of sapphire (1.31 $\mu$m)
\cite{Budriunas2015}. Such SC was used to seed the OPCPA system
driven by diode-pumped Yb:KGW and Nd:YAG lasers, finally producing
carrier envelope phase-stable sub-9 fs pulses with 5.5 TW peak
power at 1 kHz repetition rate \cite{Budriunas2017}. In the range
of anomalous GVD of sapphire, the SC spectrum spanning wavelengths
from 470 nm to more than 2.5 $\mu$m was reported with 15 fs pulses
at 2 $\mu$m, however, showing a noticeable drop of the spectral
intensity around 1 $\mu$m \cite{Darginavicius2013}.

Unlike many other laser host crystals, YAG has no birefringence
due to its cubic structure. This property as combined with
excellent crystalline quality, high nonlinearity and high optical
damage threshold makes YAG a very attractive nonlinear medium for
SC generation. As pumped in the visible spectral range (515 nm),
YAG produces a narrower ($390-625$ nm) and more structured SC
spectrum than that obtained in sapphire \cite{Bradler2014b}.
However, the advantages of YAG crystal show-up with longer pump
wavelengths. With near infrared pump wavelengths, YAG produces
quite similar SC spectrum as sapphire ($420-1600$ nm, with 800 nm
pumping), which exhibits higher spectral intensity in the
infrared, as illustrated in Fig.~\ref{fig:laserhost}(a), and which
is obtained with reduced, sub-$\mu$J pump energies due to large
nonlinear index of refraction of the crystal \cite{Bradler2009}.
The SC generation in YAG was also studied with $1.1-1.6~\mu$m
tunable pulses from an optical parametric amplifier, demonstrating
fairly stable blue cut-off at 530 nm and progressive extension of
the infrared part of the spectrum while increasing the pump
wavelength \cite{Bradler2009}.

\begin{figure}[ht!]
\includegraphics[width=8.5cm]{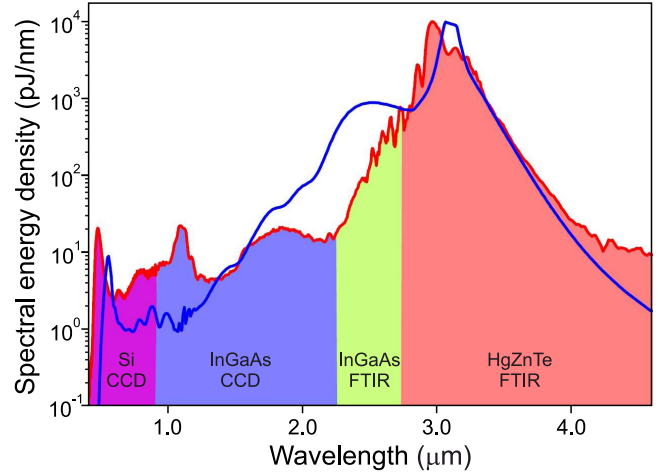}
\caption{Supercontinuum spectrum generated by 3.1 $\mu$m, 2.6
$\mu$J pulses in 2 mm-thick YAG plate (red curve). The ranges of
each spectrometer/detector (Si, InGaAs and HgCdZnTe) are indicated
by color fills. Superimposed is the angle-integrated spectrum from
the numerical simulation (blue curve). Adapted from
\cite{Silva2012}.}\label{fig:yag3}
\end{figure}

With the mid-infrared pumping, in the range of anomalous GVD (for
the input wavelengths longer than 1.6 $\mu$m), YAG shows an
advantage over sapphire in producing a much flatter SC spectrum
over the entire wavelength range. SC generation with a spectrum
extending from 510 nm to more than 2.5 $\mu$m was reported with
carrier envelope phase-stable 15 fs pulses at 2 $\mu$m
\cite{Darginavicius2013} and from 450 nm to more than 2.5 $\mu$m
with 32 fs pulses at 2.15 $\mu$m \cite{Fattahi2016}, both
demonstrating preservation of a stable carrier envelope phase of
the broadband radiation. However, in these experiments the longest
detectable wavelength was limited to 2.5 $\mu$m. The full spectral
extent of the SC was measured more recently, reporting the
generation of almost four optical octave-spanning SC with a
continuous wavelength coverage from 350 nm to 3.8 $\mu$m, as
generated using 100 fs, 2.3 $\mu$m pump pulses from an optical
parametric amplifier \cite{Garejev2017}. With even longer, 3.1
$\mu$m pump pulses, as delivered by high repetition rate OPCPA
system, the multioctave (from 450 nm to more than 4.5 $\mu$m,
Fig.~\ref{fig:yag3}), carrier envelope phase-stable SC was
reported, potentially yielding a single optical cycle
self-compressed pulses at the output of the crystal, as predicted
by numerical simulations \cite{Silva2012}.

The high optical damage threshold of sapphire and YAG in
particular, identify them as suitable nonlinear materials for SC
generation with longer laser pulses, whose durations vary from a
few hundreds of femtoseconds to a few picoseconds. The renewed
interest in SC generation in solid state dielectric media with
picosecond laser pulses was prompted by the development of novel
Yb-based solid-state lasers operating around $1~\mu$m and
dedicated to the development of compact, efficient, and
inexpensive tabletop OPCPA systems. In that regard, picosecond SC
is expected to provide a robust broadband seeding source, which
would markedly simplify the OPCPA architecture, excluding the need
for optically or electronically synchronized broadband laser
oscillator source.

The first studies aiming at stable SC generation were performed
with Ti:sapphire laser pulses of variable duration, as accessed
via either spectral narrowing or pulse chirping
\cite{Bradler2010,Bradler2011}. These investigations were
followed-up by using genuine sub-picosecond pulses from Yb:KYW
\cite{Calendron2015} and 1-ps pulses from Nd:glass
\cite{Grazuleviciute2015b,Galinis2015} lasers, demonstrating that
under carefully chosen experimental conditions, a stable and
reproducible SC is generated without damaging the nonlinear
material. Experimental and numerical investigations revealed that
filamentation of picosecond laser pulses form a more complex
spatiotemporal structure of SC, which is governed by the free
electron plasma, in contrast to femtosecond filamentation
dynamics, which is mainly driven by the multiphoton absorption
\cite{Galinis2015}. However, the spectral-temporal analysis has
shown that the red-shifted and blue-shifted spectral broadenings
are associated just with two distinct sub-pulses, which are well
separated in time. High temporal coherence of sub-picosecond
pulse-generated SC was directly verified by the spectral phase
measurements of the parametrically amplified SC pulses
\cite{Calendron2015}. Indeed, the availability of picosecond
pulse-generated SC with well-behaved spectral phase that is
compressible down to transform limit, paved new avenues in the
development of picosecond-laser pumped noncollinear optical
parametric amplifiers \cite{Kasmi2015} and of a whole new
generation of compact SC-seeded OPCPA systems. These systems are
built around amplified solely sub-picosecond and picosecond
lasers, such as Yb:KYW \cite{Emons2010}, Yb:YAG
\cite{Schulz2011,Riedel2014,Fattahi2016,Thire2017,Indra2017},
Yb-fiber \cite{Rigaud2016,Archipovaite2017}, and are designed to
provide few optical cycle pulses in the near- and mid-infrared at
very high repetition rates. Finally, picosecond SC serves as a
seed signal in the mid-infrared optical parametric amplifiers
driven by amplified picosecond Ho:YAG lasers
\cite{Kanai2017,Malevich2017}.

The SC generation was also studied in other laser host materials,
such as potassium gadolinium tungstate (KGW), gadolinium vanadate
(GdVO$_4$) and yttrium vanadate (YVO$_4$) using Ti:sapphire laser
pulses \cite{Bradler2009}. These crystals exhibit high nonlinear
refractive indexes, which result in very low critical power for
self-focusing and allow SC generation with sub-100 nJ input pulse
energies. However, due to relatively small bandgaps and
transmission cut-off wavelengths located in the near ultraviolet,
the measured spectral broadening in these materials was rather
modest, and covered the spectral ranges of $500-1150$ nm in KGW,
and $550-1150$ nm in GdVO$_4$ and YVO$_4$ crystals, see
Fig.~\ref{fig:laserhost}(b). SC generation with $800-1500$ nm
tunable pump pulses was also studied in gadolinium orthosilicate
(GSO), gallium gadolinium garnet (GGG), lithium tantalate (LTO)
and lutetium vanadate (LVO) crystals \cite{Ryba2014}. The authors
measured just the antistokes (blue-shifted) portions of the SC
spectra and recoded fairly constant cut-off wavelengths of 450 nm
in GSO and GGG, and 550 nm and 650 nm in LTO and LVO crystals,
respectively.

\subsection{Birefringent crystals}

Noncentrosymmetric nonlinear crystals which exhibit birefringence
and possess second-order nonlinearity are indispensable nonlinear
media serving for laser wavelength conversion via nonresonant
three wave interactions, such as second harmonic, sum and
difference frequency generation, and optical parametric
amplification. The joint contribution of quadratic and cubic
nonlinearities adds unique features to the SC generation process
in these crystals.

The most simple and straightforward approach of SC generation in
noncentrosymmetric crystals neglects the second-order nonlinear
effects. To date, filamentation and SC generation in potassium
dihydrogen phosphate (KDP) crystal by launching the pump beam
along the optical axis of the crystal was investigated
experimentally \cite{Kumar2008a,Kumar2008b,Yu2011} and numerically
\cite{Rolle2014}. The SC generation was also reported in the
absence of phase matching in lithium triborate (LBO)
\cite{Faccio2005} and lithium niobate (LN) \cite{Wang2013}
crystals, and in $\alpha$-barium borate ($\alpha$-BBO) crystal,
which exhibits birefringence, but vanishing second-order
nonlinearity \cite{Vasa2015}.

Filamentation under condition of phase-matched second-order
nonlinear effects has led to SC generation, which is accompanied
by simultaneous wavelength-tunable second harmonic or sum
frequency generation, as experimentally observed in basic
nonlinear crystals, such as KDP \cite{Srinivas2005,Kumar2007} and
$\beta$-barium borate ($\beta$-BBO) \cite{Ali2010,Wang2011}.

\begin{figure}[ht!]
\includegraphics[width=8.5cm]{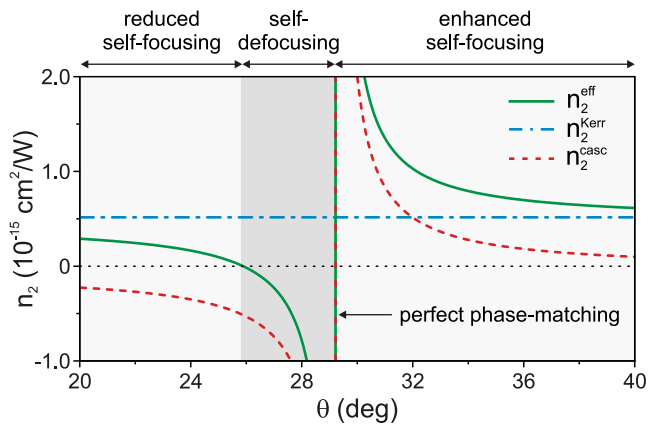}
\caption{Nonlinear refractive indices of BBO crystal at 800 nm:
intrinsic, $n_2^{\rm Kerr}$ (blue dash-dotted line), cascaded,
$n_2^{\rm casc}$ (red dashed curve) and effective, $n_2^{\rm
eff}$=$n_2^{\rm Kerr}$+$n_2^{\rm casc}$ (green solid curve) versus
the angle $\theta$, which is the angle between the propagation
direction and the optical axis of the crystal.} \label{fig:n2eff}
\end{figure}

A unique aspect of self-action phenomena in birefringent media,
which possess both, quadratic and cubic nonlinearities, and which
was generally neglected in the above studies, is the so-called
second-order cascading. The cascading effect arises from the phase
mismatched second-harmonic generation, which leads to recurrent
energy exchange between the fundamental and second harmonic
frequencies, imprinting large nonlinear phase shifts on the
interacting waves \cite{Stegeman1996}. The cascading effect mimics
the Kerr-like behavior arising from the intrinsic cubic
nonlinearity and hence produces large cascaded nonlinear index of
refraction, whose sign and magnitude could be easily varied by
rotating the crystal in the phase matching plane or changing its
temperature (Fig.~\ref{fig:n2eff}) and which is an important asset
to the nonlinear dynamics of femtosecond pulses
\cite{Conforti2011,Conforti2013}.

There are two different ways how the second-order cascading may be
favorably exploited for generation of broadband spectra in bulk
nonlinear crystals. The first approach makes use of the
self-defocusing propagation regime, where the cascaded nonlinear
index of refraction, $n_2^{\rm casc}$ is negative and its absolute
value is greater than the competing intrinsic $n_2^{\rm Kerr}$, so
$n_2^{\rm eff}<0$. Under these operating conditions, spectral
broadening is achieved without the onset of filamentation and
results from pulse self-compression and temporal soliton
generation, which in turn emerges from the opposite action of
self-phase modulation and material dispersion \cite{Zhou2012}.

\begin{figure}[ht!]
\includegraphics[width=8.5cm]{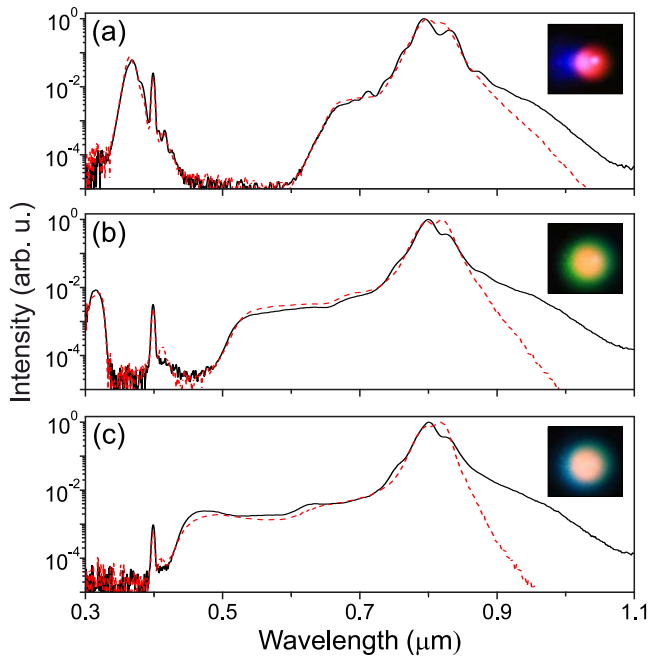}
\caption{Supercontinuum spectra in 5 mm-thick BBO crystal as
generated with 100 fs, 290 nJ input pulses at 800 nm (black solid
curves) and measured at various crystal detunings from the perfect
phase matching angle ($29.2^\circ$): (a) $32.4^\circ$, (b)
$37.5^\circ$, (c) $42.6^\circ$. The narrow peak at 400 nm and the
broad peak with tunable wavelength are the phase-mismatched and
self-phase-matched second harmonics. The insets show the visual
appearances of the SC beam in the far field. Red dashed curves
show the supercontinuum spectra generated with lower (190 nJ)
input pulse energy. Adapted from the data of
\cite{Suminas2017}.}\label{fig:bbo}
\end{figure}

The numerical simulations predicted that the soliton
compression-induced SC generation regime could be achieved in a
variety of nonlinear crystals, which possess normal GVD for the
pump wavelength \cite{Bache2013a}. More recent numerical and
experimental results show that the long wave side of the SC
radiation could be enriched by the generation of broadband
dispersive waves located beyond the zero dispersion wavelength of
the crystals: together with the soliton the dispersive waves
contribute to the octave-spanning SC, which extends from 1.0 to
4.0 $\mu$m in LN \cite{Zhou2015a} and from 0.9 to 2.3 $\mu$m in
$\beta$-BBO \cite{Zhou2015b} crystals. More recently, this
approach was extended to the nonlinear crystals that are
transparent in the mid-infrared \cite{Zhou2016}. To this end, the
SC covering wavelength range from 1.6 to 7.0 $\mu$m was
experimentally measured in lithium thioindate (LiInS$_2$) crystal,
using the pump pulses in the $3-4~\mu$m range. The soliton
compression and SC generation in self-defocusing regime is
particularly attractive concerning the energy scaling, since
favorable interplay between self-phase modulation and material
dispersion is achieved without the onset of beam filamentation.

In contrast, the second approach makes use of self-focusing
propagation regime, which leads to filamentation. Here the
cascaded nonlinearity could be used either to enhance or reduce
the effective nonlinear index of refraction, $n_2^{\rm eff}$, see
Fig.~\ref{fig:n2eff}, hence opening the possibility to perform the
nonlinear interaction in a controlled way. To this end, the
interplay between the cascaded-quadratic and intrinsic cubic
nonlinearities was favorably exploited to achieve filamentation
and SC generation with nearly monochromatic, 30 ps pulses from
Nd:YAG laser in periodically poled LN crystal, also demonstrating
control of SC spectral extent by tuning the temperature
\cite{Krupa2015}. In the femtosecond regime, filamentation and SC
generation in $\beta$-BBO crystal with the input wavelengths
falling in the range of normal \cite{Suminas2017} and anomalous
\cite{Suminas2016} GVD was reported. In the regime of normal GVD,
SC with spectral extent from 410 nm to 1.1 $\mu$m was generated
with Ti:sapphire laser pulses at 800 nm, demonstrating control of
the blue-shifted portion of the SC spectrum by tuning the angle
between the incident laser beam and the optical axis of the
crystal \cite{Suminas2017}, see Fig.~\ref{fig:bbo}. The achieved
spectral control was very robust in terms of input pulse energy
and was attributed to efficient generation of the
self-phase-matched second harmonic, which introduced a
considerable energy loss and distortion of the trailing sub-pulse
shape, counteracting the joint effect of cascaded-quadratic and
cubic self-steepenings. A broader SC spectrum, extending from 520
nm to 2.5 $\mu$m was generated with 1.8 $\mu$m pulses from an
optical parametric amplifier, whose wavelength falls into the
anomalous GVD range of the $\beta$-BBO crystal \cite{Suminas2016}.
Here the ultrabroadband SC spectrum was generated due to the
formation of spatiotemporal light bullets, which experience more
than 4-fold self-compression from 90 fs down to 20 fs. Moreover,
in the range of enhanced self-focusing (where $n_2^{\rm casc}$ is
positive), a markedly reduced filamentation threshold was
experimentally detected, that led to SC generation at sub-critical
powers for self-focusing.

The second-order cascading-mediated spectral broadening was also
reported under different experimental settings. Generation of
intense SC with a spectrum between 1.2 and 3.5 $\mu$m and with
sub-mJ energy input pulses at 1.5 $\mu$m was demonstrated in
highly nonlinear organic DAST crystal \cite{Vicario2015}. A
dramatic spectral broadening of sub-two optical cycle pulses with
a central wavelength of 790 nm was achieved in $\beta$-BBO crystal
due to multi-step cascaded interactions yielding an ultrabroadband
spectrum in the $0.5-2.4~\mu$m range \cite{Kessel2016}.

\section{Supercontinuum generation in semiconductors}

Bulk semiconductors possess much larger cubic nonlinearities than
dielectrics and extended transparency windows in the mid-infrared
spectral range, and so emerge as promising nonlinear media for
hosting the nonlinear interactions in the near- and mid-infrared
\cite{Chin2001}. The mid-infrared spectral range is particularly
interesting for spectroscopic studies since it covers the
so-called molecular fingerprint region. The feasibility of
semiconductor materials for SC generation in the mid-infrared
spectral range was experimentally demonstrated more than 30 years
ago \cite{Corkum1985}. Here the authors studied spectral
broadening in gallium arsenide (GaAs), zinc selenide (ZnSe), and
cadmium sulphide (CdS) crystals with picosecond pulses at 9.3
$\mu$m from an amplified CO$_2$ laser. The broadest SC spectrum,
which spanned from 3 to 14 $\mu$m was measured in GaAs crystal.

From the time perspective, the real progress in the field was
achieved only quite recently. $2-20~\mu$m spanning picosecond SC
in GaAs was reported using 2.5 ps, 9.3 $\mu$m pulses from CO$_2$
laser \cite{Pigeon2014}. In the femtosecond regime, the output
spectrum spanning wavelengths from 3.5 to 7 $\mu$m was generated
in a 10-mm thick GaAs crystal when pumped by 100 fs pulses with a
central wavelength of 5 $\mu$m, as obtained by the difference
frequency generation between the signal and idler pulses of the
Ti:sapphire laser-pumped near-infrared optical parametric
amplifier \cite{Ashihara2009}. More recently, a GaAs crystal was
used in a similar setup to generate the SC in the $4-9~\mu$m
range, using $4.2-6.8~\mu$m tunable pulses. The authors also
reported pulse compression down to sub-two optical cycle widths
around 6 $\mu$m by employing external post compression of
spectrally broadened pulses in BaF$_2$, CaF$_2$ and MgF$_2$
crystals, which possess anomalous GVD \cite{Lanin2014}.
$3-18~\mu$m spanning SC was generated in GaAs crystal when pumped
by femtosecond laser pulses with a central wavelength of 7.9
$\mu$m, which falls into the range of anomalous GVD of GaAs
\cite{Lanin2015}. Along with spectral superbroadening,
simultaneous pulse self-compression down to almost a single
optical cycle was measured. More than three octave-wide
supercontinuum, with the spectrum from 500 nm to 4.5 $\mu$m was
generated in a zinc sulphide (ZnS) crystal pumped with 27 fs
pulses at 2.1 $\mu$m from the OPCPA system \cite{Liang2015}.

A series of interesting results on spectral broadening and SC
generation were recently reported in a ZnSe crystal. Filamentation
and SC generation was studied in a ZnSe crystal using pump pulses
with tunable wavelength and so accessing self-focusing regimes of
different orders of multiphoton absorption \cite{Durand2014}.
However, no filamentation was observed with 800 nm pump pulses,
while only very modest spectral broadening in the near-infrared
spectral range was achieved with pump wavelength at 1.2 $\mu$m.
The potential of ZnSe crystal to produce an ultrabroadband SC for
spectroscopic applications was demonstrated using the pump pulses
with central wavelength of 5 $\mu$m, that is close to the zero GVD
wavelength (4.8 $\mu$m) of the crystal. A SC spectrum spanning
wavelengths from 500 nm to 11 $\mu$m was measured in the
multifilamentation regime \cite{Mouawad2016}.

\begin{figure}[ht!]\center
\includegraphics[width=8.5cm]{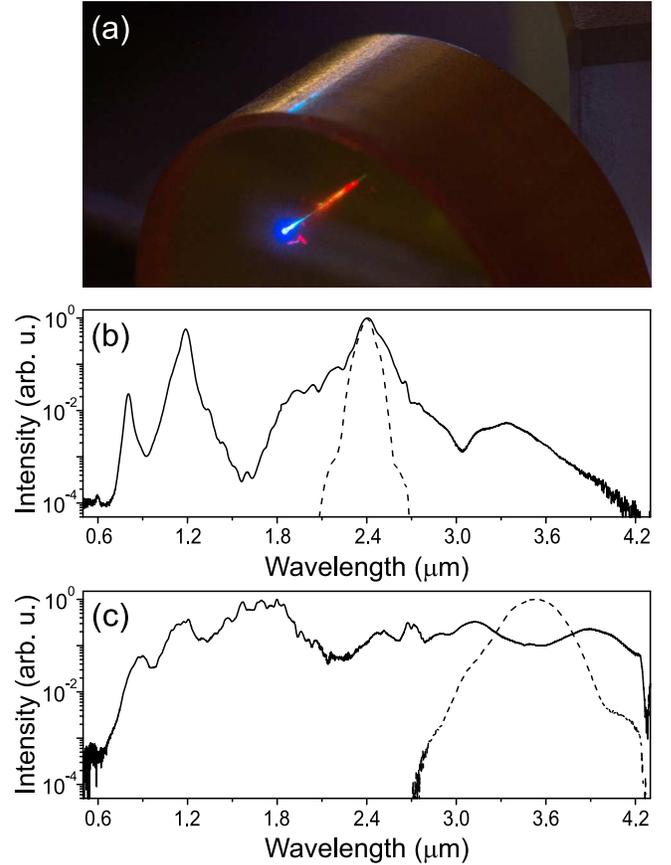}
\caption{(a) Photograph of a light filament in ZnSe excited by 100
fs, 2.4 $\mu$m pulses: the blue track is the plasma luminescence,
the red track is the visible portion of spectrally broadened third
harmonic at 800 nm. Supercontinuum spectra in 5 mm-thick
polycrystalline ZnSe plate, as generated by filamentation of: (b)
100 fs, 3 $\mu$J pulses at 2.4 $\mu$m, adapted from
\cite{Suminas2017b}, (c) 60 fs, 6.5 $\mu$J pulses at 3.5 $\mu$m.
The sharp drop of the spectral intensity at 4.25 $\mu$m is due to
CO$_2$ absorption. The input pulse spectra are shown by dashed
curves.}\label{fig:znse}
\end{figure}

In addition to its high cubic nonlinearity, ZnSe owes an
attractive set of optical properties, such as polycrystalline
structure and non-vanishing second-order nonlinearity, which
produce a number of interesting nonlinear effects, enriching the
spectral content of SC. Spectral broadening around the carrier
wavelength, accompanied by simultaneous generation of multiple
broadband even and odd harmonics was reported in polycrytalline
ZnSe samples using few optical cycle pulses at 3 $\mu$m
\cite{Archipovaite2017}, and at 3.5 $\mu$m and 5.2 $\mu$m
\cite{Kanai2017} from recently developed mid-infrared OPCPA
systems. Figure~\ref{fig:znse}(a) shows a photograph of a light
filament in 5 mm-long ZnSe sample, excited by 100 fs, 2.4 $\mu$m
pulses from an optical parametric amplifier. An ultrabroadband SC
spectrum covering the wavelength range from 600 nm to 4.2 $\mu$m,
that corresponds to 2.8 optical octaves \cite{Suminas2017b}, is
illustrated in Fig.~\ref{fig:znse}(b). Prominent broadband
spectral peaks at 1.2 $\mu$m and 800 nm are associated with second
and third harmonics, respectively, which were simultaneously
generated by three-wave mixing processes. A very efficient
harmonics generation was attributed to randomly quasi
phase-matched broadband three-wave mixing, which stems from the
disorder of the nonlinear domains (randomly oriented individual
crystallites) \cite{Baudrier2004}. Self-focusing and filamentation
of longer-wavelength (3.5 $\mu$m) input pulses, generated by the
difference frequency mixing between the signal and idler waves of
the optical parametric amplifier, produced a reasonably flat,
plateau-like SC spectrum, which results from merging of the
spectral broadening around the incident wavelength and the second,
third and fourth harmonics peaks, extending beyond the long-wave
detection limit of the spectrometer, as shown in
Fig.~\ref{fig:znse}(c),

\section{Control of supercontinuum generation}

The rapidly expanding field of applications calls for achieving
broadband radiation with desired temporal and spectral properties,
which in turn require setting an efficient control on the
filamentation process. As described in the previous sections, the
filamentation dynamics and hence the spectral content of the SC
are defined essentially by the laser wavelength and by linear and
nonlinear properties of the medium, such as material dispersion,
bandgap and the nonlinear index of refraction
\cite{Brodeur1998,Kolesik2003a}. These material parameters possess
fundamental mutual relationships \cite{Sheik-Bahae1990} and
therefore are generally fixed for a given nonlinear medium at a
given input wavelength, so reducing the possibilities to influence
the filamentation and SC generation dynamics in real experimental
settings.

Several conceptually different approaches have been proposed to
overcome these limitations, allowing to modify the spectral
content of the SC radiation in a controlled fashion. The first
approach relies on tailoring and shaping the parameters of the
input laser beam and pulse. To this end, a certain control of
filamentation and SC generation processes was achieved by varying
the polarization state (from linear to circular) of the incident
radiation in isotropic media \cite{Sandhu2000,Srivastava2003} and
by choosing either ordinary or extraordinary input pulse
polarizations in birefringent media with zero second-order
susceptibility, e.g. $\alpha$-BBO crystal \cite{Vasa2015}. Phase
and amplitude shaping of the input pulse allowed to significantly
(by the order of magnitude) increase the spectral intensity of the
SC within specified bandwidths \cite{Schumacher2002}. More
recently, tailoring the input pulse shape by introducing second-
and third-order phase distortions which were controlled by an
acousto-optic programmable dispersive filter, was demonstrated to
affect the pulse splitting dynamics, which in turn resulted in
generation of the SC with controllable spectral bandwidth and
shape \cite{Dharmadhikari2015}. A similar approach based on
employing a 4-f pulse shaper with a phase mask produced by liquid
crystal spatial light modulator was demonstrated for achieving
accumulation of the spectral intensity at selected frequencies by
a controllable amount \cite{Thompson2017}.

\begin{figure}[ht!]
\includegraphics[width=8.5cm]{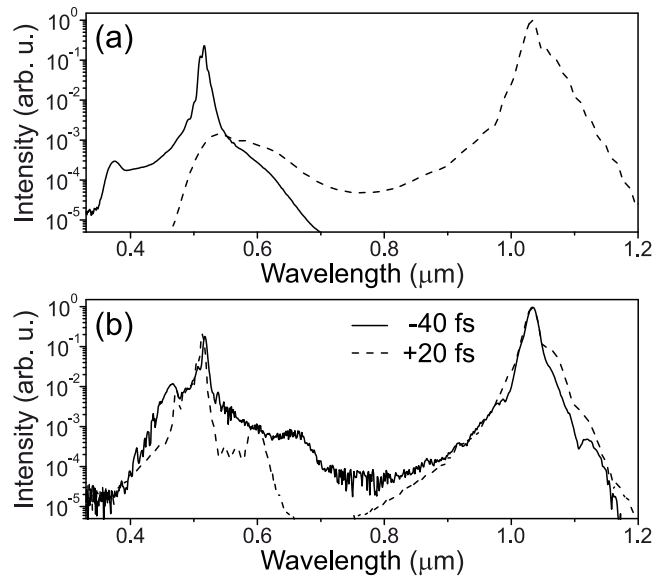}
\caption{(a) Individual supercontinuum spectra generated in 3
mm-long sapphire plate by filamentation of 300 fs pulses at 1030
nm and 515 nm, each containing a power equal to 2.6 $P_{\rm cr}$.
(b) Resulting SC spectra generated by co-filamentation of the
above pulses. The negative delay time indicates that the second
harmonic pulse comes first. Courtesy of M.
Vengris.}\label{fig:cofil}
\end{figure}

Setting of either positive or negative chirp \cite{Kartazaev2007}
or adjustment of the focal plane of the input pulses
\cite{Faccio2008} allowed controllable tuning of the blue-shifted
cut-off of the SC spectrum. On the other hand, enhancement of the
red-shifted SC spectral extent, was achieved by setting low
numerical aperture of the input beam and using a longer nonlinear
medium \cite{Bradler2009,Jukna2014}. A notable extension of the SC
spectrum and enhancement of the SC components within certain
spectral ranges in particular, was achieved by means of two color
filamentation, i.e. by launching two filament-forming beams of
different wavelengths \cite{Wang2006}. An additional degree of
freedom in controlling the SC generation process via the time
delay between the input pulses was demonstrated using crossing
beams in the single \cite{Kolomenskii2016} and to some extent in
the multiple \cite{Stelmaszczyk2009} filamentation regimes.
Figure~\ref{fig:cofil}(a) shows the individual SC spectra in
sapphire, generated by filamentation of separately launched 300 fs
pulses at fundamental (1030 nm) and second harmonic (515 nm)
wavelengths from an amplified Yb:KGW laser.
Figure~\ref{fig:cofil}(b) shows an example how the spectral
intensity within certain spectral regions and the overall shape of
the resulting SC spectrum is modified by varying the time delay
between the co-filamenting fundamental and second harmonic pulses.
An efficient control of the SC spectrum was also performed by
changing the relative position of the focus in the nonlinear
medium by means of diffractive optics
\cite{Romero2011,Borrego2013} or by varying the spatial phase of
the input beam with spatial light modulators
\cite{Kaya2012,Borrego2014}. Fine adjustment of the position of
the nonlinear focus was demonstrated by varying the carrier
envelope phase of a few-cycle input pulse \cite{Zhong2014}.

The second approach is based on tailoring the linear (absorption)
and nonlinear properties (nonlinear index of refraction) of the
medium itself. As suggested in the pioneering work
\cite{Jimbo1987}, various dopants may lead to an enhanced
nonlinear optical response of the medium and thus lowering the
threshold condition for nonlinear processes, which in turn result
in a modification of SC spectral shape. At present, these early
ideas received further attention by demonstrating SC generation in
silver \cite{Wang2007} and gold \cite{Vasa2013} nanoparticles and
gold nanorods \cite{Vasa2014} doped water and in LN crystal doped
with Ni ions \cite{Wang2013}. Femtosecond SC generation was also
reported in more complex nonlinear media such as aqueous colloids
containing silver nanoparticles \cite{Driben2009} and
nanocomposite materials \cite{Kulchin2013}. Filamentation of
femtosecond laser pulses in aqueous solutions with absorbing
inorganic \cite{Wang2010} or protein \cite{Santhosh2010} dopants
was shown to produce more spectrally flat SC. In contrast,
narrowed and discontinuous SC spectra were generated in
laser-modified nonlinear solid-state media, such as fused silica
\cite{Paipulas2012} and YAG \cite{Grazuleviciute2015}. Very
promising results regarding efficient control of the SC spectrum
were obtained by making use of the second-order cascading effects
in birefringent crystals, as discussed in more detail in Section
5E of the paper.

\section{Power scaling and applications to pulse compression}

The emerging applications in many fields of ultrafast science and
extreme nonlinear optics require high power broadband radiation
and few optical cycle pulses that are generated by all-solid-state
technology and by relatively simple means. Pulse compression based
on increasing the spectral bandwidth via self-phase modulation in
a bulk solid-state medium and subsequent removal of the frequency
modulation by using an appropriate dispersive delay line,
represents a simple and universal method for obtaining few optical
cycle pulses \cite{Rolland1988,Mevel2003}.

\begin{figure}[ht!]
\includegraphics[width=8.5cm]{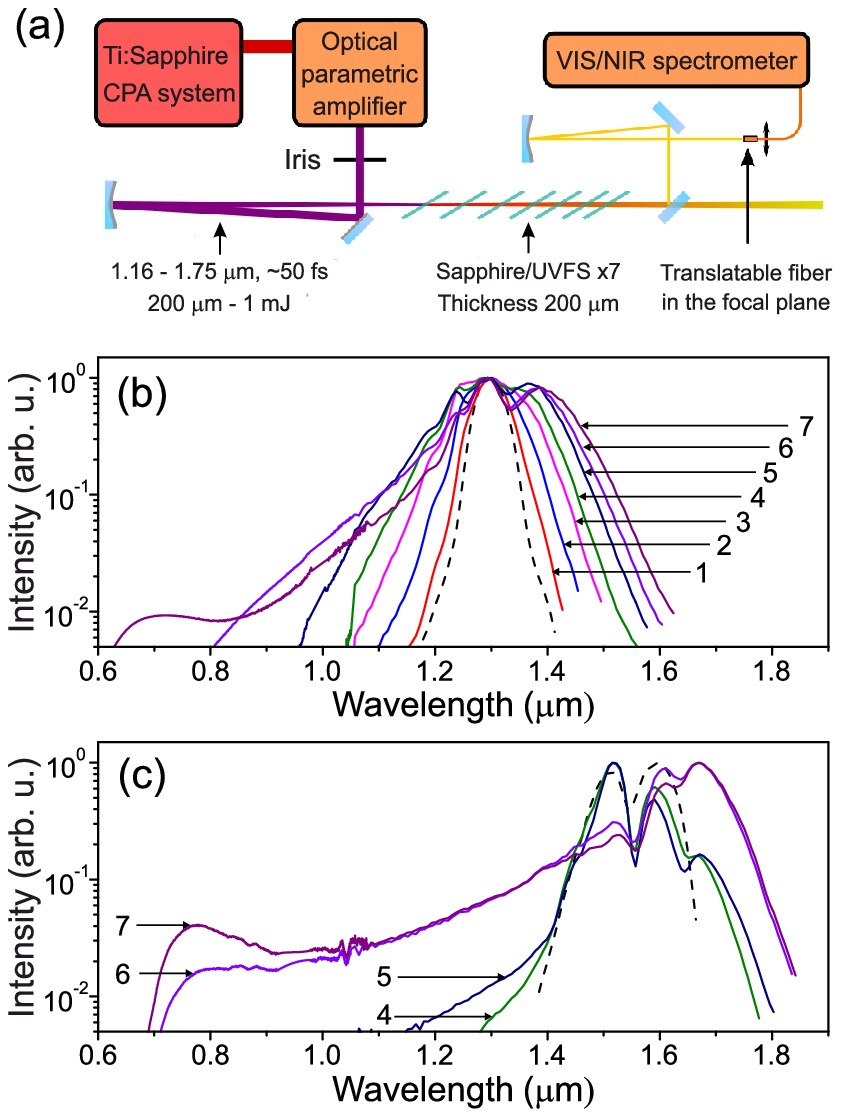}
\caption{(a) Experimental setup for supercontinuum generation in
multiple alternating plates of UV-grade fused silica and sapphire.
Supercontinuum spectra as functions of the plate number as
recorded with the input pulse wavelengths of (b) 1.3 $\mu$m, (c)
1.6 $\mu$m, which fall into the ranges of zero and anomalous GVD,
respectively. The input spectra are shown by dashed curves. The
input pulses energies were 0.36 mJ at 1.3 $\mu$m and 0.5 mJ at 1.6
$\mu$m. Adapted from \cite{Budriunas2017b}.}\label{fig:plates}
\end{figure}

Recently, a flexible and versatile approach for the generation of
high power, high energy SC in a solid-state medium was proposed
and demonstrated by using a series of distributed thin plates
instead of a single piece of bulk nonlinear medium \cite{Lu2014}.
The individual fused silica plates were located so that the
optical pulse exits each plate before the onset of small scale
self-focusing and break-up of high power beam into multiple
filaments, yielding high energy broadband radiation with a uniform
spatial profile. The proposed approach was further elaborated by
demonstrating the scaling of the input peak power to of as much as
two thousand times the critical power for self-focusing in the
solid-state medium \cite{Cheng2016}. Remarkably, the cascaded
spectral broadening in a distributed multi-plate arrangement was
demonstrated to yield the SC pulse with a regular chirp, which is
compressible to a nearly transform-limit using external pulse
compressor \cite{Lu2014}, and this approach could be easily
implemented in many schemes for achieving few optical cycle
pulses. Indeed, 250 fs output pulses from Yb:YAG thin disc
oscillator were compressed down to 15 fs \cite{Seidel2016}. More
recently, the input 0.8 mJ, 30 fs laser pulses at 790 nm were
spectrally broadened in a sequence of 7 fused silica plates of 100
$\mu$m thickness each and compressed using chirped mirrors down to
5.4 fs (two optical cycles) with almost 90\% throughput efficiency
\cite{He2017}. A multi-plate approach was also used to generate
high power broadband SC in a sequence of 7 alternating UV-grade
fused silica and sapphire plates of 200 $\mu$m thickness
[Fig.~\ref{fig:plates}(a)] with tunable pump wavelength from 1.1
to 1.75 $\mu$m, which covers the regimes of normal, zero and
anomalous GVD of fused silica and sapphire and with pump energy up
to 0.5 mJ \cite{Budriunas2017b}. Figures~\ref{fig:plates}(b) and
(c) show the SC spectra as functions of the plate number as
recorded with the input pulse wavelengths of 1.3 $\mu$m and 1.6
$\mu$m, respectively. A comparison between the SC generated in
multiple plates and continuous nonlinear medium showed that the
energy scaling is achieved without the degradation of essential
performance characteristics, such as shot-to-shot energy stability
and stability of the carrier envelope phase of the output pulses.
Finally, the numerical simulations predict that multiple thin
plates of solid-state medium separated by appropriate spatial
filters could be employed to broaden the spectrum of ultrahigh
peak power pulses, yielding spectral bandwidths that are
compressible to few-cycle pulsewidths, with undistorted output
beam profiles that are focusable to ultrarelativistic intensities
\cite{Voronin2013}.

An even more simple and thus very attractive approach for
extracavity pulse compression makes use of the spectral broadening
in bulk solid-state media featuring anomalous GVD. In contrast,
here no additional dispersive element to compensate the self-phase
modulation is required: the job is done by the anomalous GVD of
the medium itself. The numerical simulations suggest that
mid-infrared pulses may be self-compressed down to a single
optical cycle limit \cite{Silva2012}. To this end, generation of
filament-induced spatiotemporal light bullets, which are
self-compressed down to or even below two optical cycles was
experimentally reported in various bulk solid-state dielectric
media
\cite{Smetanina2013,Chekalin2015,Liang2015,Grazuleviciute2015,Suminas2016,Grazuleviciute2016},
as noted in the previous sections of the paper.

\begin{figure}[ht!]
\includegraphics[width=8.5cm]{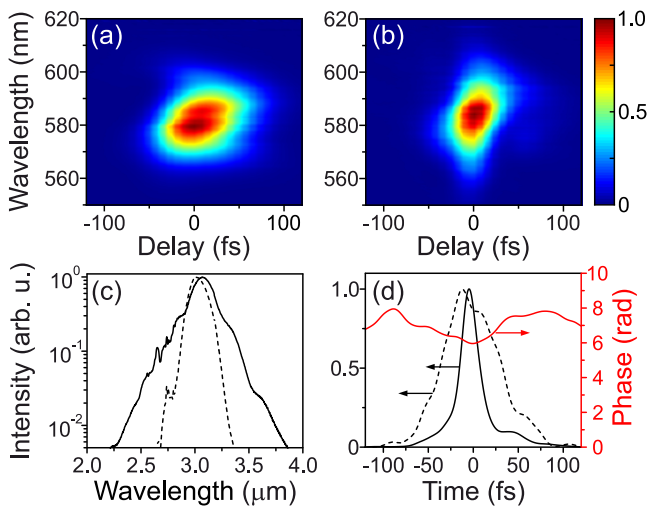}
\caption{Self-compression of 70 fs, $3~\mu$m pulses in 3 mm-thick
YAG plate. Measured SFG-FROG traces of (a) input, (b)
self-compressed pulses, (c) spectrum and (d) retrieved intensity
profile and phase of the self-compressed pulse. Dashed curves in
(c) and (d) show the spectrum and retrieved intensity profile of
the input pulse. Adapted from
\cite{Marcinkeviciute2017}.}\label{fig:compress}
\end{figure}

However, fully evolved self-compressed light bullets contain just
a relatively small fraction of the input energy and exhibit
dispersive broadening as they leave the nonlinear medium and
propagate in free space \cite{Majus2014}, and these features may
appear undesirable for a range of practical applications.
Therefore a more attractive realization of the anomalous
GVD-induced self-compression mechanism relies upon using a shorter
nonlinear medium and larger input beam, extracting the
self-compressed pulse before the filamentation regime sets in. In
that way, sub-3 optical cycle self-compressed pulses with
$>10~\mu$J energy were obtained at 3 $\mu$m by placing 3 mm-thick
YAG plate at the output of high repetition rate OPCPA system
\cite{Hemmer2013,Baudisch2015}. Self-compression of parametrically
amplified difference frequency pulses in few mm-thick YAG, CaF$_2$
and BaF$_2$ crystals down to sub-3 optical cycle widths was
demonstrated in the $3-4~\mu$m spectral range
\cite{Marcinkeviciute2017}. Figure~\ref{fig:compress} presents the
experimental results illustrating a three-fold self-compression of
70 fs pulses with the carrier wavelength of $3~\mu$m down to 23 fs
(2.3 optical cycles) in a 3 mm-thick YAG plate, due to the
interplay between the self-phase modulation and anomalous GVD and
without the onset of self-focusing effects. An energy throughput
efficiency of above 90\% was measured, yielding the
self-compressed pulses with sub-$30~\mu$J energy, with the only
energy losses occurring due to Fresnel reflections from the input
and output faces of the nonlinear crystal.

Prospects for wavelength and power scaling of the self-compression
technique were investigated numerically
\cite{Bravy2014,Bravy2015,Li2017}, suggesting that scaling to
milijoule energies and terawatt peak powers is readily feasible as
long as one dimensional dynamics of the input pulse are maintained
and small-scale filamentation of large beams is avoided
\cite{Voronin2016a,Voronin2016b}. More recently, the numerical
predictions were verified experimentally, where sub-three-cycle
(30 fs) multimilijoule pulses with 0.44 TW peak power were
compressed in YAG crystal of 2 mm thickness and extracted before
the onset of modulation instability and multiple filamentation, as
a result of favorable interplay between strong anomalous GVD and
optical nonlinearity around the carrier wavelength of 3.9 $\mu$m
\cite{Shumakova2016}.

\section{Other developments}

A different mechanism of pulse compression and spectral broadening
was uncovered for the propagation of intense (few TW/cm$^2$)
mid-infrared pulses in a thin piece of nonlinear medium ($\sim2$
mm CaF$_2$ plate) \cite{Garejev2016}. Here the self-compression of
the driving pulse was shown to occur due to generation of free
electron plasma, which expels the trailing part of the pulse out
of propagation axis. The plasma-induced pulse compression resulted
in spectral broadening around the carrier wavelength as well as
facilitated large scale spectral broadening of third, fifth and
seventh harmonics via cross phase modulation, eventually yielding
more than 4 octave-spanning SC from the ultraviolet to the
mid-infrared. Interestingly, the overlapping spectra of odd
harmonics produce a remarkable spectral extension into the deep
ultraviolet, which is well beyond the blue-shifted cut-off
constrained by the material GVD \cite{Kolesik2003a}.
Plasma-induced compression along with broadband four-wave mixing
was shown to dramatically enhance the SC generation in GaAs around
its zero-GVD wavelength (6.8 $\mu$m) \cite{Stepanov2016}.
Post-compression of an octave-spanning spectrum in 0.5-mm BaF$_2$
plate yielded 20 fs pulse at the output, that is less than 0.9
field cycles.

Only a moderate broadening of the SC spectrum was observed with
Ti:sapphire laser pumping in narrow bandgap dielectric crystals,
such as diamond \cite{Kardas2013}. However, recent numerical
simulations predict that filamentation of femtosecond mid-infrared
pulses in medium and narrow bandgap dielectric media, which
possess extremely broad mid-infrared transmittance and whose zero
GVD points are located deeply in the mid-infrared, might produce
the multi-octave SC spectra with remarkable red-shifted spectral
broadening \cite{Frolov2016}. For instance, the simulated SC
spectra in sodium chloride and potassium iodide using the pump
wavelength of 5 $\mu$m, cover wavelength ranges of $0.7-7.6~\mu$m
and $0.66-22~\mu$m, respectively, suggesting these media as
potentially attractive alternatives to soft glasses and
semiconductor crystals. An interesting experimental study on SC
generation with 1.24 $\mu$m femtosecond pulses from Cr:forsterite
laser was performed in various aggregate states of CO$_2$:
high-pressure gas, liquid, and supercritical fluid
\cite{Mareev2016}. The authors reported the SC generation in the
$0.4-2.2~\mu$m range, whose spectral shape was strongly dependent
on the pressure.

So far, the vast majority of experiments on SC generation were
performed using conventional, Gaussian-shaped input beams.
However, several studies were devoted to investigate the SC
generation with more complex input waveforms. In that regard, SC
generation was studied using femtosecond Bessel beams with various
parameters: cone angle, central spot size, energy content,
particular intensity variation along the propagation axis and
length of the Bessel zone. Self-focusing of ultrashort-pulsed
Bessel beams was investigated in various nonlinear media, such as
water \cite{Dubietis2007}, methanol \cite{Sun2012}, sapphire
\cite{Majus2013} and BaF$_2$ \cite{Dota2012,Dota2014} and
demonstrated that the SC radiation appears on the propagation
axis. However, due to energy distribution between the central spot
and surrounding rings, Bessel beams are very resistant to
self-focusing, and 10-100 times higher incident energy is required
to induce filamentation and achieve appreciable spectral
broadening. Filamentation and SC generation with self-bending Airy
beams in water \cite{Polynkin2009,Ament2012} and in fused silica
\cite{Gong2016}, has shown that the patterns of SC emission
provide quantitative clues on the complex evolution of these
sophisticated waveforms in the highly nonlinear propagation
regime. Finally, the SC generation was investigated in CaF$_2$
with vortex \cite{Neshev2010} and singular \cite{Maleshkov2011}
beams, which are associated with the presence of a spiral and
step-wise phase dislocations in the wavefront of a beam,
respectively, and which determine its phase and intensity
structure. In both cases it was shown that the generated SC
experiences large divergence and appears as a wide white-light
background surrounding the original input beam, whose spatial
profile remains well preserved in the process of SC generation.

\section{Conclusions}

To summarize, the present state of art of ultrafast SC generation
in bulk condensed media has reached a high level of maturity
thanks to a wealth of new exciting experimental results, which
were backed-up by in-depth understanding of femtosecond
filamentation phenomena. The advances in SC generation in bulk
solid-state media boosted the development of all-solid-state laser
technology, which made high power few optical cycle pulses
accessible throughout the optical range. Such pulses attract great
scientific and technological interest and will continue to do so
in the future, extending the ultrafast nonlinear optics and laser
spectroscopy to new, previously inaccessible domains. In
particular, SC generation represents one of the fundamental
building blocks of the emerging so-called third-generation
femtosecond technology, which foresees boosting the peak and
average powers of few optical cycle pulses simultaneously to the
multiterawatt and hundreds of watts range, respectively, paving
the way for the generation of powerful sub-cycle pulses with full
control over the generated light waves in the optical range and
beyond \cite{Fattahi2014}.

\section*{Acknowledgments}

We are particularly indebted to prof. E. Riedle, prof. E. Mazur,
prof. M. Vengris, dr. A. Varanavi\v{c}ius and R. Budri\={u}nas for
their kind permissions to reproduce some of their data. We also
thank N. Garejev and A. Marcinkevi\v{c}i\={u}t\.{e} for their help
in performing the measurements dedicated to illustrate important
issues discussed throughout the paper. This work was funded by the
Research Council of Lithuania, grant APP-8/2016.

\end{document}